\documentclass[reprint,amsmath,amssymb,aps,superscriptaddress, longbibliography]{revtex4-1}
\usepackage{amssymb}
\usepackage{graphicx}
\usepackage{graphics}
\usepackage{color}
\usepackage{enumerate}

\begin{document}

\title{Enhanced phase estimation with coherently-boosted two-mode squeezed beams and its application to optical gyroscopes}

\author{Xiao-Qi Xiao}
\email[]{xiaoxq@sdju.edu.cn}
\affiliation{Department of Communication Engineering,
Shanghai Dianji University, Shanghai 200240, China}
\affiliation{Hearne Institute for Theoretical Physics and Department of Physics and Astronomy, Louisiana State University, Baton Rouge, Louisiana 70803, USA}

\author{Elisha S. Matekole}
%\email[]{esiddi1@lsu.edu}
\affiliation{Hearne Institute for Theoretical Physics and Department of Physics and Astronomy, Louisiana State University, Baton Rouge, Louisiana 70803, USA}

\author{Jiankang Zhao}
%\email[]{zhaojiankang@sjtu.edu.cn}

\author{Guihua Zeng}
%\email[]{ ghzeng@sjtu.edu.cn}
\affiliation{State Key Laboratory of Advanced Optical Communication Systems and Networks,\\ Institute of Sensing and Navigation,\\ Department of Electronic Engineering, Shanghai Jiaotong University,\\ Shanghai 200030, China}

\author{Jonathan P. Dowling}
%\footnote{Deceased.}
\email[]{Deceased June 5, 2020}
\affiliation{Hearne Institute for Theoretical Physics and Department of Physics and Astronomy, Louisiana State University, Baton Rouge, Louisiana 70803, USA}
\affiliation{National Institute of Information and Communications Technology,
Tokyo 184-8795, Japan}
\affiliation{NYU-ECNU Institute of Physics at NYU Shanghai, Shanghai 200062,
China}
\affiliation{CAS-Alibaba Quantum Computing Laboratory, USTC, Shanghai
201315, China}

\author{Hwang Lee}
%\email[]{hwlee@lsu.edu}
\affiliation{Hearne Institute for Theoretical Physics and Department of Physics and Astronomy, Louisiana State University, Baton Rouge, Louisiana 70803, USA}

\date{\today}

\begin{abstract}
Quantum techniques, developed in recent decades, provide new approaches to achieving high-precision measurements beyond the classical bounds. In this paper, we theoretically demonstrate a metrology method for improving the sensitivity of the interferometric optical gyroscope, robust against the loss,  by using coherent-light stimulated two-mode squeezed beams as the light source. The detection protocol is based on a simple intensity measurement, and the quantum noise is far below the shot-noise limit. The enhancement factors for different coherent light fields are analyzed in detail. Additionally, the influence of loss during the propagation in the optical path is studied, and the conditions for achieving sub-shot-noise measurement sensitivity are obtained. We also find that the phase sensitivity of the proposed gyroscope scheme becomes closer to the quantum Cram\'{e}r-Rao bound with increasing of the photon number of the coherent beams.
\end{abstract}

\pacs{}

\maketitle

\section{Introduction}

The advent of the laser in 1960 evoked interest in the use of the Sagnac effect for sensing inertial rotation by optical means. Generally, in an optical gyroscope, there is a pair of counter-propagating light beams traveling around a ring cavity, and the interference between them provides information about the rotation rate relative to an inertial frame. To be specific, the phase difference $\varphi $ between the two counter-propagating beams after one round trip is given by $\varphi =8\pi A\Omega /c\lambda $, where $\Omega $ is the angular velocity of the rotating cavity, $\lambda $ is the wavelength, and $A$ is the area enclosed by the light beams. The theoretical limit on the minimum detectable angular velocity $\Omega _{\min}$ is directly related to the uncertainty in measuring $\varphi $, given by \cite{chow1985ring}
\begin{equation}
\Omega _{\min}=\frac{c\lambda }{8\pi A}\Delta \varphi .
\label{a}
\end{equation}
The sensitivity of the gyroscope is then determined by the phase resolution of the interferometers.

In general, the Sagnac effect has always used lasers, in spite of a few with atoms or single photons \cite{simmonds2001quantum,stedman1997ring,bertocchi2006single,fink2019entanglement}, and the minimum detectable phase shift is the shot-noise limit for $\Delta \varphi_{\rm{SN}} = 1/\sqrt{\left\langle N\right\rangle }$, where $\left\langle N\right\rangle $ is the average number of photons.
Recent theoretical and experimental work has proved that the advance in quantum techniques makes it possible to enhance the phase-sensitivity scale of an interferometer beyond the conventional bound; even approaching or beyond the Heisenberg limit, $\Delta \varphi _{\rm{HL}} = 1/\left\langle N\right\rangle $ \cite{fink2019entanglement,caves1981quantum,yurke1986input,ou1997fundamental,holland1993interferometric,dowling1998correlated,kolkiran2007heisenberg,yurke19862,hofmann2009all,anisimov2010quantum}. In particular, the squeezing light technique has gained much attention in improving measurement sensitivity of the interferometer \cite{anisimov2010quantum,birrittella2015coherently,kolkiran2008quantum, steuernagel2004approaching,kimble2001conversion,olivares2007optimized,sparaciari2015bounds,sparaciari2016gaussian}. The Mach-Zehnder interferometer with two independently-squeezed coherent states has been investigated with respect to quantum Fisher information (QIF) and the detection loss effect on phase sensitivity \cite{olivares2007optimized,sparaciari2015bounds,sparaciari2016gaussian}. The two-mode squeezed state, which has a correlation feature between the two modes, has also been used as the input light source to enhance the phase sensitivity with parity measurements \cite{anisimov2010quantum, birrittella2015coherently}, coincidence measurements \cite{kolkiran2008quantum}, and homodyne detection \cite{steuernagel2004approaching}.

The improvement in accuracy implies a revolution in the field of interferometric sensing. In terms of the atomic gyroscope, in 1998 Dowling demonstrated a type of gyroscope based on an entangled two-mode Fock state, the sensitivity of which can achieve the Heisenberg limit \cite{dowling1998correlated}. Recently, Heisenberg-limited Sagnac interferometry with entangled-photon pairs from parametric down conversion has been reported by using high-order coincidence measurements \cite{kolkiran2007heisenberg}. In addition, a nonlinear Sagnac interferometer is proposed by replacing the beam splitter (BS) with a nonlinear optical process \cite{xin2017nonlinear}. Although these works showed a great enhancement in phase sensitivity, the difficulty in producing the light source, or implementing the detection strategy, or even modifying the structure of the Sagnac interferometer, would undoubtedly put an obstacle in their way toward application with current technology. 

In this paper, we investigate the sensitivity of the optical gyroscope fed with a two-mode squeezed-light field generated from an optical parametric amplification (OPA) process with two-mode coherent light field inputs---the so-called coherent-boosted two-mode squeezed beams. The phase shift is estimated by measuring the intensity difference of the light beams directly, which is much simpler than the multi-photon coincidence measurement \cite{kolkiran2007heisenberg, kolkiran2008quantum}  or parity measurement \cite{anisimov2010quantum, birrittella2015coherently} suggested in previous works. The sub-shot-noise  sensitivity with robustness against the loss is the main focus of our discussion. The enhancement factor with respect to the shot-noise limit is studied in detail under the conditions that the two modes of the coherent field are balanced or unbalanced in modulus, as well as in phase. We note that the measurement accuracy has seldom been examined in the latter condition.
In addition, since loss plays an important role in precision measurements, especially in the fiber gyroscope, we explore the influence of the loss arising from the light propagation. We analyze the requirements on the loss and reciprocity of the two paths for sub-shot-noise measurements. 
Moreover, we study the quantum Cram\'{e}r-Rao bound for our gyroscope scheme to evaluate the efficiency  of the intensity difference estimation strategy. It should be pointed out that the results obtained in this paper can be applied to other kinds of SU(2) interferometers such as the Mach-Zehnder interferometer \cite{yurke19862}.

Unlike the conventional optical gyroscope, the biased phase modulation is no longer necessary for the gyroscope discussed in the present paper, and thus the corresponding loss can be avoided. Meanwhile, the loss arising from the propagation in the gyroscope can be overcome by control of the OPA parameters.
All of  this means that our gyroscope scheme has advantages not only in measurement  precision but also in robustness. On the other hand, compared with interferometric schemes with a two-mode squeezed vacuum state  (TMSV) \cite{steuernagel2004approaching}, or one squeezed input and one coherent input \cite{kimble2001conversion}, our scheme contains more photons experiencing the phase shift. This is due to the coherent light field stimulation in the OPA, and thus a higher sensitivity can be reached. The measurement sensitivity of the proposed gyroscope is very close to that of the coherent boosted SU(1,1) interferometer \cite{plick2010coherent}, or the nonlinear interferometer \cite{ou2012enhancement}. At the same time, the tolerance to loss in the interferometer is better than the nonlinear interferometer in Ref. \cite{ou2012enhancement}.

The paper is organized as follows. In Sec. \ref{og}, our gyroscope scheme for phase measurement beyond the shot-noise limit is described. The influence of the loss due to propagation is investigated in Sec. \ref{noise}. In Sec. \ref{qcrb}, the effect of the detection strategy is discussed. Then, in Sec. \ref{discussion} we compare the present gyroscope scheme with the traditional one and the previous works. The conclusion is given in Sec. \ref{conclusion}.

\begin{figure}[h]
\centering\includegraphics[width=8.3cm]{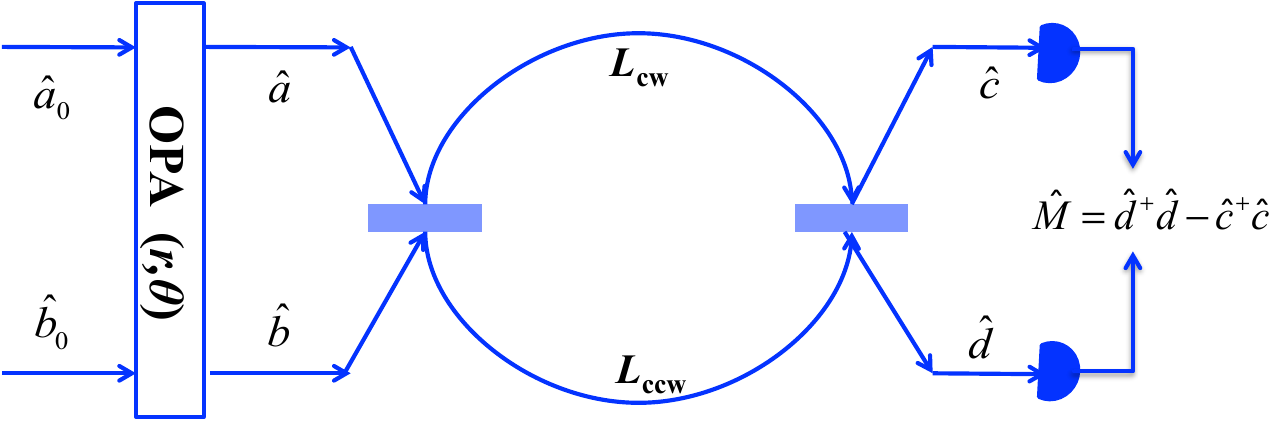}
\caption{The equivalent optical network diagram of the Sagnac interferometer for the coherent-boosted two-mode squeezed light field, where the first and second beam splitters are meant to be one and the same in the Sagnac interferometer. The  two-mode squeezed light fields in $\hat{a}$ and $\hat{b}$, which are generated by the OPA (with the parameters $r$ and $\theta$), are injected into the gyroscope from the beam splitter. After traveling in the clockwise and counter-clockwise directions, respectively, the two light fields combine at the beam splitter again, and then leave the gyroscope. Then, immediately, the intensity difference $\langle \hat{M}\rangle$ between the two output fields is detected to estimate the phase difference between the two arms. }
\label{1}
\end{figure}

\section{\label{og}Optical gyroscope beyond the shot-noise limit}

As described in Fig. \ref{1}, a two-mode coherent light field, corresponding to the annihilation operators $\hat{a}_{0}$ and $\hat{b}_{0}$, respectively, is seeded into an OPA pumped by a high-power laser. The operation of the OPA can be described by a  two-mode squeeze operator $\hat{S}=e^{\xi ^{*}\hat{a}\hat{b}-\xi \hat{a}^{\dag}\hat{b}^{\dag}}$ with $\xi =r e^{i \theta}$, and the field operators of the output modes $\hat{a}$ and $\hat{b}$ can be written as \begin{eqnarray}
\hat{a} &=&\hat{a}_{0}\cosh r-e^{i\theta }\hat{b}_{0}^{\dagger}\sinh r, \label{2}\\
\hat{b} &=&\hat{b}_{0}\cosh r-e^{i\theta }\hat{a}_{0}^{\dagger}\sinh r, \label{3}
\end{eqnarray}
where $r$ is the parametric gain coefficient, and $\theta $ is the phase of the pump laser \cite{scully1997quantum}.

The output light of the OPA is fed into the Sagnac interferometer, the equivalent optical path diagram of which is depicted in Fig. \ref{1}.
We assume that the BS of the gyroscope is 50-50, and the phase shift on transmission at the BSs for both the input and the output light field is $0$, and that on reflection is $\pi /2$. After passing through the BS for the first time, the light has two possible paths, one is in the clockwise direction $L_{\rm{cw}}$ and the other is counter-clockwise $L_{\rm{ccw}}$. If the interferometer is not rotating, the optical paths for both directions are the same; otherwise, the optical path difference will be induced. When the two light beams recombine at the BS for the second time, the light emerges from the two output ports and is detected by two detectors. We describe the light field leaving the interferometer by operators $\hat{c}$ and $\hat{d}$ in the following form:
\begin{eqnarray}
\hat{c} &=&\frac{1}{2}\left[ \left( e^{i\mu }-e^{i\nu }\right) \hat{a}+\left( \mathrm{i}e^{i\mu }+\mathrm{i}e^{i\nu }\right) \hat{b}\right] , \\
\hat{d} &=&\frac{1}{2}\left[ \left( \mathrm{i}e^{i\mu }+\mathrm{i}e^{i\nu }\right) \hat{a}
-\left( e^{i\mu }-e^{i\nu }\right) \hat{b}\right] ,
\end{eqnarray}
where $\mu \propto L_{\rm{cw}}$ and $\nu \propto L_{\rm{ccw}}$ are the phases proportional to the length of the two light paths \cite{scully1997quantum}.

Suppose the initial two-mode coherent light fields $\hat{a}_{0}$ and $\hat{b}_{0}$ are in the state $\vert \psi \rangle =\left\vert \alpha \right\rangle \left\vert \beta \right\rangle $. The average number of the total photons exiting the interferometer then is:
\begin{eqnarray}
\left\langle \hat{N}\right\rangle _{\mathrm{tot}} &=&\left\langle \hat{c}^{\dagger}\hat{c}+\hat{d}^{\dagger}\hat{d}\right\rangle =\left\langle \hat{a}^{\dagger}\hat{a}+\hat{b}^{\dagger}\hat{b}\right\rangle  \nonumber \\
&=&(\left\vert \alpha \right\vert ^{2}+\left\vert \beta \right\vert ^{2})\cosh 2r-(\alpha ^{\ast }\beta ^{\ast }e^{i\theta }+\alpha \beta e^{-i\theta })\sinh 2r   \nonumber\\
& &+2\sinh ^{2}r,
\label{6}
\end{eqnarray}
where the loss in the optical paths is omitted. We notice that, generally, the average photon numbers of the coherent light fields $\left\vert \alpha \right\vert^{2},\mbox{ } \left\vert \beta \right\vert ^{2}>>\sinh ^{2}r$ (this assumption is used for the rest of Secs. II and III), and then the above equation can be approximated as
\begin{eqnarray}
\left\langle \hat{N}\right\rangle _{\mathrm{tot}} &\approx &(\left\vert \alpha \right\vert ^{2}+\left\vert \beta \right\vert ^{2})\cosh 2r   \nonumber\\
& &-(\alpha ^{\ast }\beta ^{\ast }e^{i\theta }+\alpha \beta e^{-i\theta })\sinh 2r.
\label{7}
\end{eqnarray}

To find out the phase signal produced by the difference of the two light paths, we measure the intensity difference operator between the two output beams, that is,
\begin{eqnarray}
\hat{M} &=&\hat{d}^{\dag }\hat{d}-\hat{c}^{\dag }\hat{c}  \nonumber \\
&=&(\hat{a}^{\dag }\hat{a}-\hat{b}^{\dag }\hat{b})\cos \varphi +(\hat{a}^{\dag }\hat{b}+\hat{a}\hat{b}^{\dag })\sin \varphi  ,
\label{8}
\end{eqnarray}
where $\varphi =\mu -\nu$ is the phase difference between the two beams. Hence, the output signal is given by
\begin{eqnarray}
\left\langle \hat{M}\right\rangle &=&(\left\vert \alpha \right\vert ^{2}-\left\vert \beta \right\vert ^{2})\cos \varphi +\{(\alpha ^{\ast }\beta +\alpha \beta ^{\ast})\cosh 2r  \nonumber \\
& &-\frac{1}{2}[e^{\mathrm{i}\theta }(\alpha ^{\ast 2}+\beta ^{\ast 2})+e^{-\mathrm{i}\theta }(\alpha ^{2}+\beta ^{2})]\sinh 2r\}   \nonumber\\
& &\times \sin \varphi .
\label{c}
\end{eqnarray}

The uncertainty of the phase signal is given by the linear error propagation method \cite{scully1997quantum,gerry2005introductory}
\begin{equation}
\Delta \varphi ^{2}=\frac{\Delta M^{2}}{\left\vert \partial \left\langle
\hat{M}\right\rangle /\partial \varphi \right\vert ^{2}},
\label{10}
\end{equation}%
which is dependent on the phase shift $\varphi $. The linear error propagation method is conventionally used for the local phase estimation, in which a small change of signal is monitored at a fixed reference point \cite{durkin2007local}.
As for the gyroscope, we are concerned with the sensitivity at the point where the phase shift is close to zero, under the assumption that the rotation we desire to detect is usually very small.
%Moreover, since the detected signal $\left\langle \hat{M}\right\rangle$, as described in Eq. (\ref{c}), is the sine function of the phase$\varphi $, the maximal response sensitivity can be obtained at the point $\varphi =0$.
Thus, we analyze the uncertainty of the phase in the case of $\varphi =0$. We have
\begin{widetext}
\begin{equation}
\Delta \varphi ^{2}=\frac{\left\vert \alpha \right\vert ^{2}+\left\vert \beta \right\vert ^{2}}{\left\vert (\alpha ^{\ast }\beta +\alpha \beta ^{\ast })\cosh 2r-\frac{1}{2}[e^{\mathrm{i}\theta }(\alpha ^{\ast 2}+\beta ^{\ast 2})+e^{-\mathrm{i}\theta }(\alpha ^{2}+\beta ^{2})]\sinh 2r\right\vert ^{2}}.
\label{b}
\end{equation}
\end{widetext}
The uncertainty of the phase signal is not only dependent on the OPA parameters $r$ and $\theta $, but is also determined by the complex amplitudes of the initial two-mode coherent light fields $\alpha $ and $\beta $.

On the other hand, according to Eq. (\ref{7}), the shot-noise limit for the phase measurement, i.e. the variance of the phase shift for the classical gyroscope with the coherent light source containing the same average photon number, is
\begin{widetext}
\begin{equation}
\Delta \varphi _{\mathrm{SN}}^{2} =\frac{1}{\left\langle \hat{N}\right\rangle _{\mathrm{tot}}}
=\frac{1}{(\left\vert \alpha \right\vert ^{2}+\left\vert \beta \right\vert ^{2})\cosh 2r-(\alpha ^{\ast }\beta ^{\ast }e^{\mathrm{i}\theta }+\alpha ^{\ast }\beta ^{\ast }e^{-\mathrm{i}\theta })\sinh 2r}.
\end{equation}
\end{widetext}
Next, we will discuss the behavior of the phase sensitivity in the following three cases classified according to the complex amplitudes of the initial coherent light field. This will make it easy to compare the relationship between the variance of the phase shift $\Delta \varphi $ and the shot-noise limit $\Delta \varphi _{\mathrm{SN}}$. To make the comparison convenient, we define a factor $R=\Delta \varphi ^{2}/\Delta \varphi ^{2}_{\mathrm{SN}}$ to indicate the enhancement of the phase sensitivity. The uncertainty of the phase $\Delta \varphi $ will be below the shot-noise limit if $R<1$.

\begin{enumerate} [(i)]

\item Case I: $\alpha =\beta $

We first focus on the case that the complex amplitude of the input coherent light field for each mode is the same, that is, $\alpha =\beta =\eta =\sqrt{n}e^{i\gamma }$ (assuming $n$ and $\gamma $ to be real), where $n$ is the average photon number of each coherent beam. Therefore, the average total number of the photons of the initial coherent light field is $\langle \hat{N}\rangle _{\rm{coh}}=2n$. In this condition, the average number of the photons [based on Eq. (\ref{7})] exiting from the Sagnac ring is
\begin{eqnarray}
\left\langle \hat{N}\right\rangle _{\mathrm{tot}}& =& 2n\cosh 2r-2n\sinh 2r\cos (\theta -2\gamma ) \nonumber \\
& =& 2nG =\langle \hat{N}\rangle _{\rm{coh}}G.
\end{eqnarray}
where $G=\cosh 2r-\sinh 2r\cos (\theta -2\gamma )$, which is the parametric power gain of the OPA. It varies with the parametric gain coefficient $r$, the phase of the pumping laser $\theta $, and the phase of the initial coherent light field $\gamma $. The shot-noise limit is $\Delta \varphi _{\rm{SN}}^{2} =1/(2nG)$. Meanwhile, according to Eq. (\ref{b}), we can calculate the uncertainty of the phase shift as
\begin{equation}
\Delta \varphi ^{2}=\frac{1}{2nG^{2}}=\frac{1}{\langle \hat{N} \rangle_{\rm{coh}}G^{2}}=\frac{1}{G}\Delta \varphi _{\rm{SN}}^{2}, \label{e} 
\end{equation}
For $G>1$, the minimum  detectable phase shift $\Delta \varphi $ is less than the shot noise; while, for $G<1$, it is greater than the shot noise. The maximum value of $G$, given by $G_{\max }=e^{2r}$, is obtained when $\cos (\theta -2\gamma)=-1$, i. e., $\theta =\pi +2\gamma $. Hence, the variance of the phase shift will be
\begin{equation}
\Delta \varphi ^{2} =\frac{1}{2n(e^{2r})^{2}}=e^{-2r}\Delta \varphi _{\mathrm{SN}}^{2}.
\label{15}
\end{equation}
Here, we see that our gyroscope is shown to be doubly enhanced over the conventional (classical) counterpart: First, the shot noise itself is reduced by the increase of input photon number due to the parametric amplification [see Eq. (13)].  Then, there is the signal-to-noise improvement via squeezing as shown in Eq. (15).  At this stage, the enhancement factor $R$ of the phase uncertainty can be written as
\begin{equation}
R=\Delta \varphi ^{2}/\Delta \varphi ^{2}_{\mathrm{SN}}=e^{-2r}.
\label{17}
\end{equation}
The above equation shows that the uncertainty of the phase shift is decreased with the parametric gain coefficient $r$ relative to the shot-noise limit, which means the uncertainty in measuring the angular velocity $\Omega $ will be reduced accordingly.

\item{}Case II: $\left\vert \alpha \right \vert \neq \left\vert \beta \right\vert $, but $\gamma _{\alpha}=\gamma _{\beta } =\gamma $, where $\gamma _{\alpha}$ and $\gamma _{\beta}$\ are the phases of the complex amplitudes $\alpha $\ and $\beta $, respectively.

\begin{figure}[t]
\centering
\begin{minipage}[b]{\linewidth}
\centering
\includegraphics[width=3.4in]{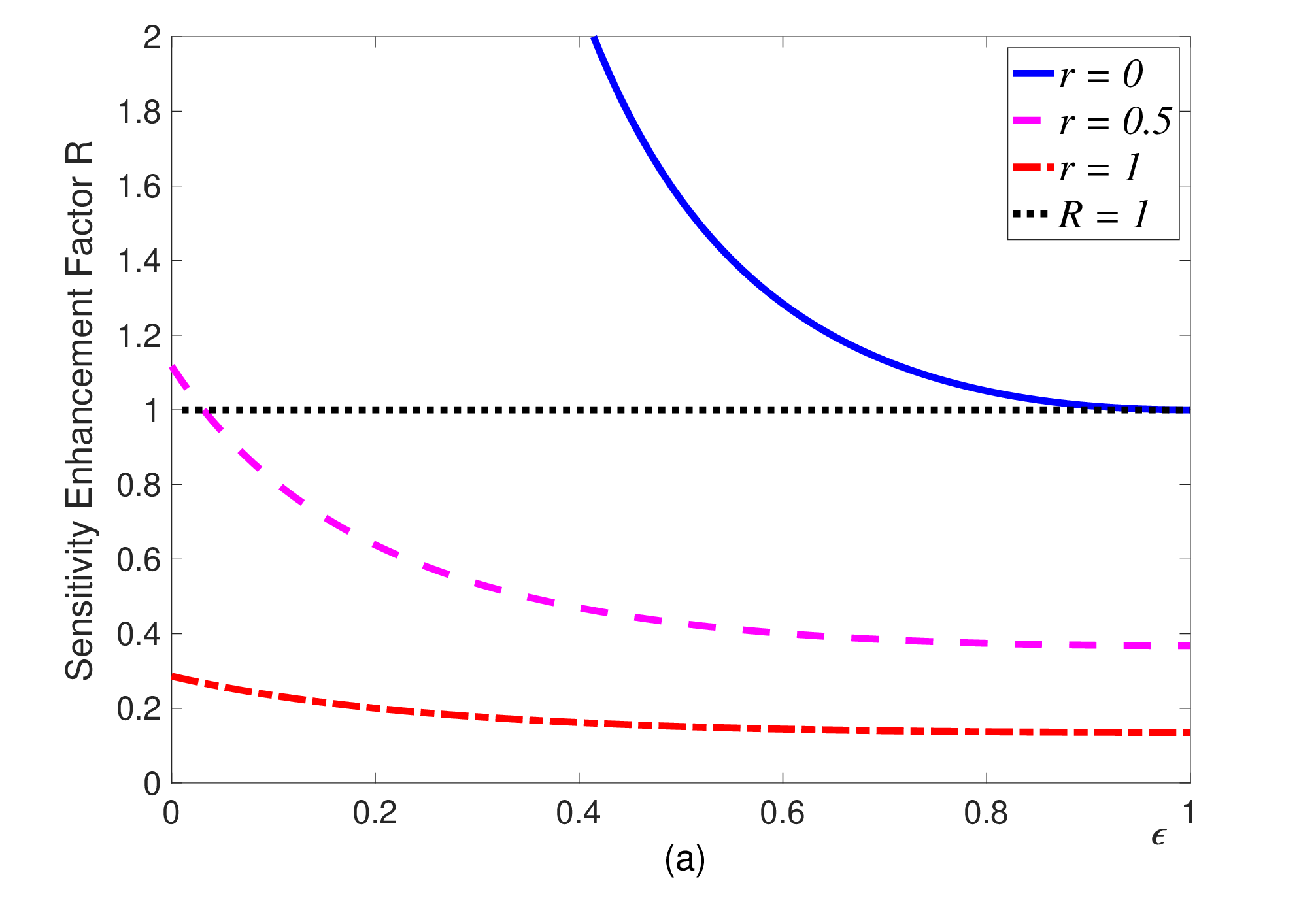}
%\caption{fig2}
\end{minipage}%
 \quad
\begin{minipage}[b]{\linewidth}
\centering
\includegraphics[width=3.4in]{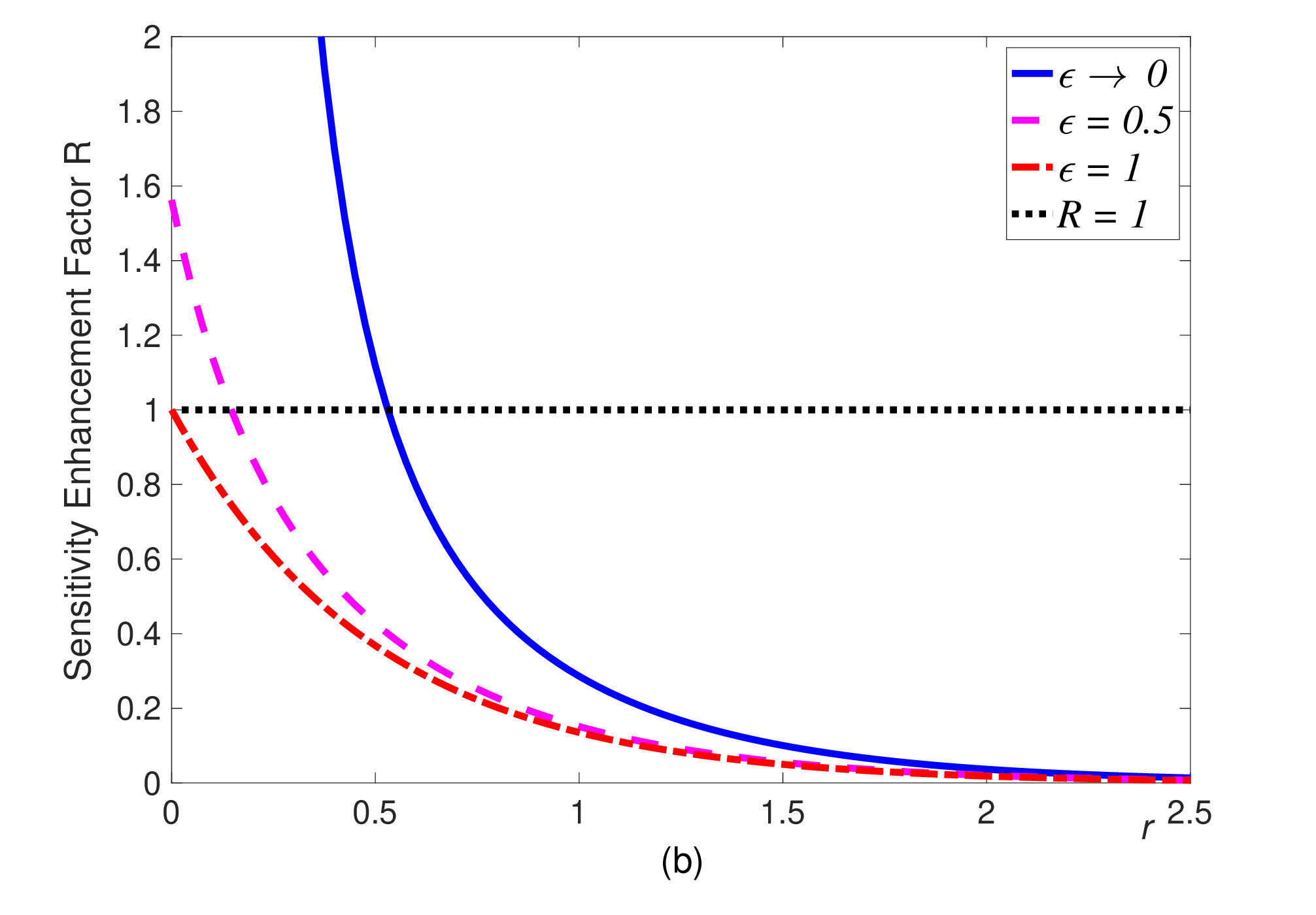}
%\caption{fig2}
\end{minipage}
\caption{For case II, the sensitivity enhancement factor $R$ between the variance of the phase $\Delta \varphi ^{2}$ and shot-noise limit $\Delta \varphi ^{2}_{\mathrm{SN}}$ varies with (a) the parameter $\epsilon $ and (b) the parametric gain coefficient $r$. $R=1$ (the black dot line) means the phase sensitivity reaches the shot-noise limit.}
\label{2}
\end{figure}

Here we consider the case that there is a difference in the intensity but not in the phase between the two amplitudes of the initial coherent light. Suppose that $\alpha =\sqrt{n}e^{\mathrm{i}\gamma }$, $\beta =\epsilon \sqrt{n}e^{\mathrm{i}\gamma }$, where $0<\epsilon \leqslant 1$ indicates the ratio between the intensity of the two modes $\hat{a}_{0}$ and $\hat{b}_{0}$, then we have $\left\vert \beta \right\vert /\left\vert \alpha \right\vert =\epsilon $. According to Eq. (\ref{b}), the variance of the phase takes the following form
\begin{equation}
\Delta \varphi ^{2}=\frac{1+\epsilon ^{2}}{n\left\vert 2\epsilon \cosh 2r-(1+\epsilon ^{2})\sinh 2r\cos (\theta
-2\gamma )\right\vert ^{2}}.
\end{equation}
Based on Eq. (\ref{7}) the shot-noise limit is
\begin{equation}
\Delta \varphi ^{2}_{\mathrm{SN}}=\frac{1}{n[(1+\epsilon ^{2})\cosh 2r-2\epsilon \sinh 2r\cos (\theta
-2\gamma )]}.
\end{equation}
Both $\Delta \varphi ^{2}$ and $\Delta \varphi ^{2}_{\mathrm{SN}}$ will each have their own minimum when the condition $\theta =\pi +2\gamma $ is satisfied, and the ratio between them is
\begin{equation}
R=\frac{(1+\epsilon ^{2})[(1+\epsilon ^{2})\cosh 2r+2\epsilon \sinh 2r]}{\left\vert 2\epsilon \cosh 2r+(1+\epsilon ^{2})\sinh 2r\right\vert ^{2}}
\end{equation}

%\begin{figure}[t]
%\centering\includegraphics[width=9cm]{pic2.eps}
%\caption{For Case II, the sensitivity enhancement factor $R$ between the variance of the phase $\Delta \varphi ^{2}$ and shot-noise limit $\Delta \varphi ^{2}_{\mathrm{SN}}$ varies with (a) the parameter $\epsilon $ and (b) the parametric gain coefficient $r$. $R=1$ (the black dot line) means the phase sensitivity reaches the shot-noise limit.}
%\label{2}
%\end{figure}

Figure \ref{2} describes the dependence of the enhancement factor $R$ on $\epsilon$ and the parametric gain coefficient $r$, respectively. When $\epsilon =1$, we have $R=e^{-2r}$,  which corresponds to case I. As shown in Fig. \ref{2}(a), with the decrease of the parameter $\epsilon $, the factor $R$ increases, which means the uncertainty of phase $\Delta \varphi $ increases and is generally close to or larger than the shot-noise limit $\Delta \varphi _{\mathrm{SN}}$. However, we notice that even in the case of $\epsilon \rightarrow0$, i.e., $\left\vert \alpha \right\vert \gg \left\vert \beta \right\vert $ ($R\to \frac{\cosh 2r}{\vert \sinh2r \vert ^2}$ in this condition), the phase variance below the shot-noise limit is also obtained if the parametric gain coefficient $r$ satisfies the condition $\cosh 2r/\sinh ^{2}2r <1$ [see Fig. \ref{2}(b)].

\item{}Case III: $\left\vert \alpha \right \vert =\left\vert \beta \right\vert =\sqrt{n}$, but $\gamma _{\alpha}\neq \gamma _{\beta}$.

Lastly, we deal with the case that the two coherent states in modes $\hat{a}_{0}$ and $\hat{b}_{0}$ have the same intensity but are different in the phase of the complex amplitude. We now assume that the phase difference between the complex amplitude of the two states $\left\vert \alpha \right\rangle$ and $\left\vert \beta \right\rangle$ is $\delta $, and $\alpha =\sqrt{n}e^{\mathrm{i}\gamma }$ and $\beta =\sqrt{n}e^{\mathrm{i}(\gamma +\delta )}$. Following Eq. (\ref{b}), it is straightforward to get that the variance of the phase
\begin{widetext}
\begin{equation}
\Delta \varphi ^{2}=\frac{2}{n\left\vert 2\cosh 2r\cos \delta -\sinh 2r[\cos (\theta -2\gamma )+\cos (\theta -2\gamma -2\delta )]\right\vert ^{2}},
\label{d}
\end{equation}
\end{widetext}
and the corresponding shot-noise limit is
\begin{equation}
\Delta \varphi ^{2}_{\mathrm{SN}}=\frac{1}{2n[\cosh 2r-\sinh 2r\cos (\theta -2\gamma -\delta )]}.
\end{equation}
The optimal OPA occurs when $\theta =\pi +2\gamma +\delta $, and in this case the factor $R$ between $\Delta \varphi ^{2}$ and $\Delta \varphi ^{2}_{\mathrm{SN}}$ takes the following form
\begin{equation}
R=\frac{1}{\cos ^{2}\delta (\cosh 2r+\sinh 2r)}=\frac{e^{-2r}}{\cos ^{2}\delta }.
\label{23}
\end{equation}

%\begin{figure}[t]
%\centering
%\includegraphics[width=9cm]{pic3.eps}
%\caption{For Case III, the ratio $R$ between the variance of the phase $\Delta \varphi ^{2}$ and shot-noise limit $\Delta \varphi ^{2}_{\mathrm{SN}}$ varies with (a) the parameter $\delta $ and (b) the parametric gain coefficient $r$ . $R=1$ (the black dot line) means the phase sensitivity reaches the shot-noise limit.}
%\label{3}
%\end{figure}

\begin{figure}[t]
\centering
\begin{minipage}[b]{\linewidth}
\centering
\includegraphics[width=3.4in]{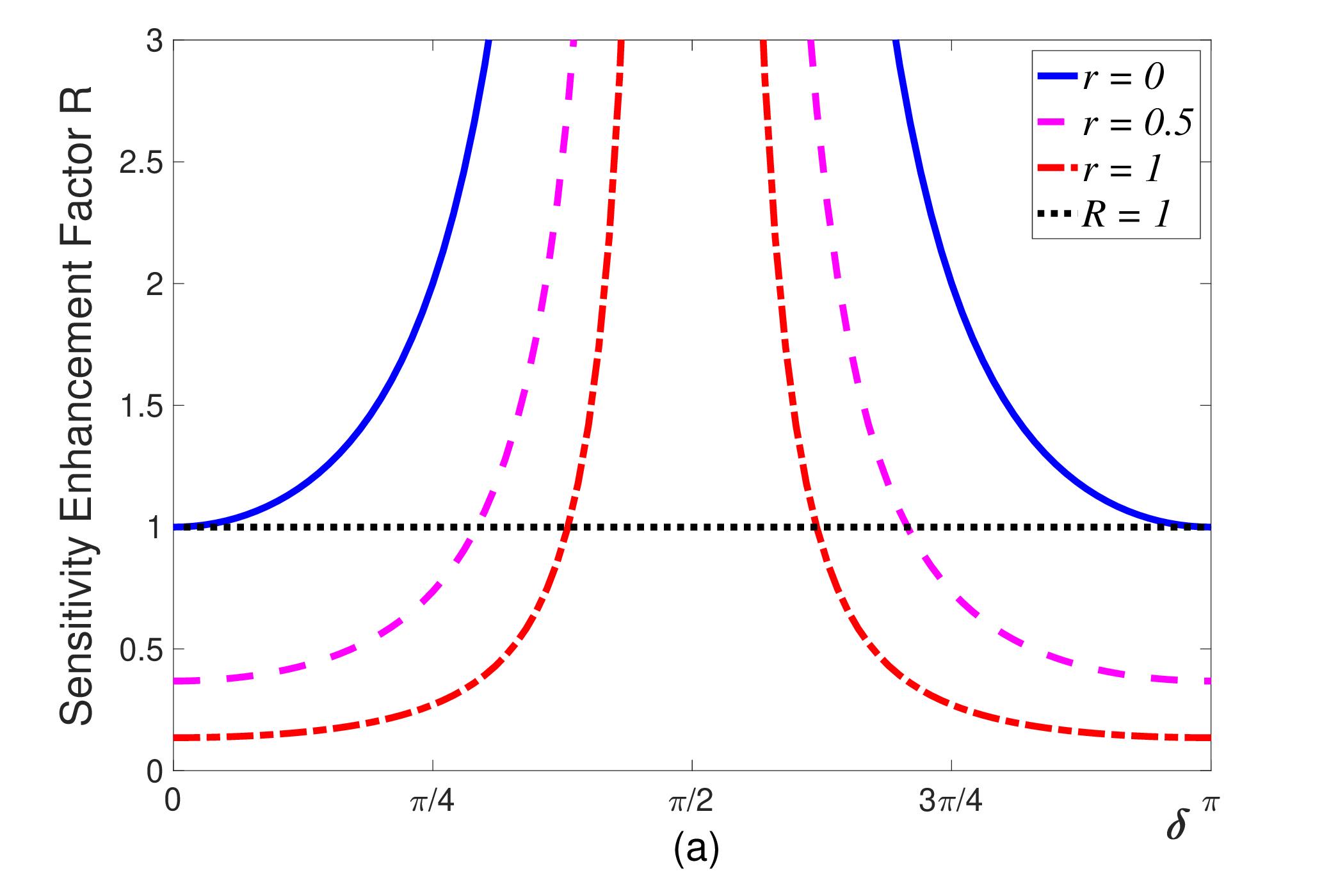}
%\caption{fig2}
\end{minipage}%
 \quad
\begin{minipage}[b]{\linewidth}
\centering
\includegraphics[width=3.4in]{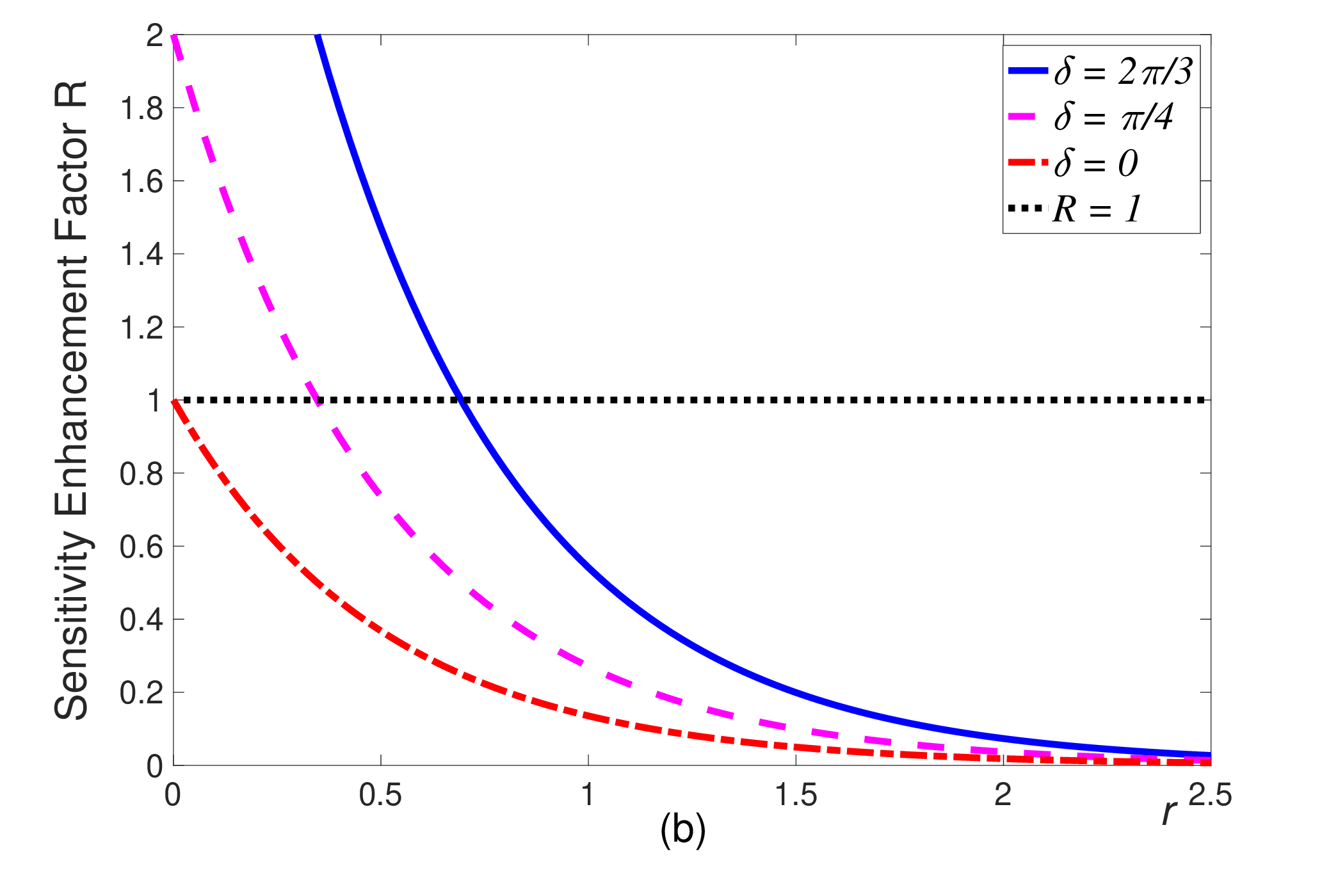}
%\caption{fig2}
\end{minipage}
\caption{For case III, the ratio $R$ between the variance of the phase $\Delta \varphi ^{2}$ and shot-noise limit $\Delta \varphi ^{2}_{\mathrm{SN}}$ varies with (a) the parameter $\delta $ and (b) the parametric gain coefficient $r$ . $R=1$ (the black dot line) means the phase sensitivity reaches the shot-noise limit.}
\label{3}
\end{figure}

The factor $R$ varies periodically with parameter $\delta $. As shown in Fig. \ref{3}(a), for one period $[0,\pi]$, $R$ increases as the parameter $\delta $ increases in the interval $[0,\pi /2)$, and decreases in the interval $(\pi /2,\pi ]$. And when $\delta \longrightarrow \pi /2$, the ratio $R\longrightarrow \infty $. This periodic variation is due to the interference between the two coherent beams with different phases. The changes of the $R$ caused by the parametric gain coefficient $r$ and the parameter $\delta $ are displayed in Fig. \ref{3}(b). The curves show that the sub-shot-noise limit phase sensitivity is not always achieved if parameter $\delta $ does not equal $0$ or $\pi $. But, for a fixed value of $\delta$ the ratio $R$ will decrease, even below one, with the increase of the parametric gain coefficient $r$.

\end{enumerate}

By comparing these three cases, it is not difficult to find that the optimal performance of the phase sensitivity takes place in the case that the input coherent light fields for the OPA are balanced (i.e. case I), in spite of the fact that the unbalanced input can also result in super-sensitivity in phase measurements. The enhancement in phase measurements can be attributed, in part, to the suppression of quantum noise of the intensity difference between the two squeezed light beams $\hat{a}$ and $\hat{b}$ at the input ports. Here, we use a new quantity, noise reduction factor  ($\mathrm{NRF}$) to indicate this suppression (see Appendix B). 

%The variance of the photon number difference between the two modes $\Delta \hat{k}=\Delta (\hat{a}^{\dagger}\hat{a}-\hat{b}^{\dagger}\hat{b})$ is suppressed below the classical limit $\Delta \hat{k}_{\mathrm{SN}}^{2} =\left\langle \hat{N}\right\rangle  _{\mathrm{tot}}$ due to the OPA procedure, and from Eqs. (2) and (3) the Noise Reduction Factor ($\rm{NRF}$) is given as
%\begin{widetext}
%\begin{eqnarray}
%\mathrm{NRF}=\frac{\Delta \hat{k}^{2}}{\Delta \hat{k}_{\mathrm{SN}}^{2}} &=& \frac{\langle (\hat{a}^{\dagger}\hat{a}-\hat{b}^{\dagger}\hat{b})^{2}\rangle -\langle \hat{a}^{\dagger}\hat{a}-\hat{b}^{\dagger}\hat{b}\rangle ^{2}}{\left\langle \hat{N}\right\rangle  _{\mathrm{tot}}} \nonumber\\
%&=&\frac{\left\vert \alpha \right\vert ^{2}+\left\vert \beta \right\vert ^{2}}{[(\left\vert \alpha \right\vert ^{2}+\left\vert \beta \right\vert ^{2})\cosh 2r-2\left\vert \alpha \right\vert \left\vert \beta \right\vert \sinh 2r\cos (\theta -\gamma _{\alpha }-\gamma _{\beta })]}
%\end{eqnarray}
%\end{widetext}
In case I with $\theta =\pi +2\gamma $, there is $\mathrm{NRF}=G^{-1}=e^{-2r}$ [see Eq.(\ref{B5})] in Appendix B], which is exactly equal to the phase sensitivity enhancement factor $R$. For case II, we have [see Eq. (\ref{B6}) in Appendix B]
\begin{equation}\mathrm{NRF}=\frac{1+\epsilon ^{2}}{(1+\epsilon ^{2})\cosh 2r+2\epsilon \sinh 2r},
\end{equation}
which increases with the decreasing of the parameter $\epsilon $. In other words, the increasing imbalance between the intensity of the two states $\left\vert \alpha \right\rangle $ and $\left\vert \beta \right\rangle$ will diminish the noise reduction, and at the same time, the phase sensitivity of the gyroscope becomes worse.

Note, however, the phase sensitivity enhancement of the gyroscope is not only determined by the noise suppression of the intensity difference of the light source of the gyroscope. As for case III, when the condition $\theta =\pi +2\gamma +\delta $ is satisfied, the $\rm{NRF}$ for the two-mode squeezed light beams will also be $e^{-2r}$ [see Eq.(\ref{B7}) in Appendix B]. Although the factor $\rm{NRF}$ is the same as that in case I,  the variance of the phase is not the same [see Eqs. (\ref{17}) and (\ref{23})]. Notice that the complex amplitude of the two squeezed coherent light beams $\hat{a}$ and $\hat{b}$ are given as follows:
\begin{eqnarray}
\left\langle \hat{a}\right\rangle &=&\sqrt{n}(e^{\mathrm{i}\gamma }\cosh r-e^{\mathrm{i}(\theta -\gamma -\delta)}\sinh r)=Ae^{\mathrm{i}\gamma _{a}},\\
\langle \hat{b}\rangle &=&\sqrt{n}(e^{\mathrm{i}(\gamma +\delta )}\cosh r-e^{\mathrm{i}(\theta -\gamma )}\sinh r)=Be^{\mathrm{i}\gamma _{b}}.
\end{eqnarray}
The phase of these two light beams then takes form
\begin{eqnarray}
\gamma _{a}&=&\arctan \frac{\cosh r\cos \gamma _{\alpha}-\sinh r\cos (\theta -\gamma _{\beta } )}{\cosh r\sin \gamma _{\alpha}-\sinh r\sin (\theta -\gamma _{\beta} )} ,\label{A}\\
\gamma _{b} &=&\arctan \frac{\cosh r\cos \gamma _{\beta }-\sinh r\cos (\theta -\gamma _{\alpha })}{\cosh r\sin \gamma _{\beta }-\sinh r\sin (\theta -\gamma _{\alpha })} .\label{B}
\end{eqnarray}
It is clear that $\gamma _{a}$ and $\gamma _{b}$ are equal to each other in cases I and II, but are not the same in case III. In particular, when $\delta =\pi /2$, there is  $\tan \gamma _{a}\tan \gamma _{b}=-1$, i.e., the phases of the two beams are separated by $\pi/2$, and the uncertainty of the phase approaches infinity (Fig. \ref{3}). Obviously, for case III, an additional interference happens between two light beams with different phases, which makes the phase sensitivity worse.

\section{Influence of loss in the optical path}
\label{noise}

\begin{figure}[t]
\centering\includegraphics[width=8.3cm]{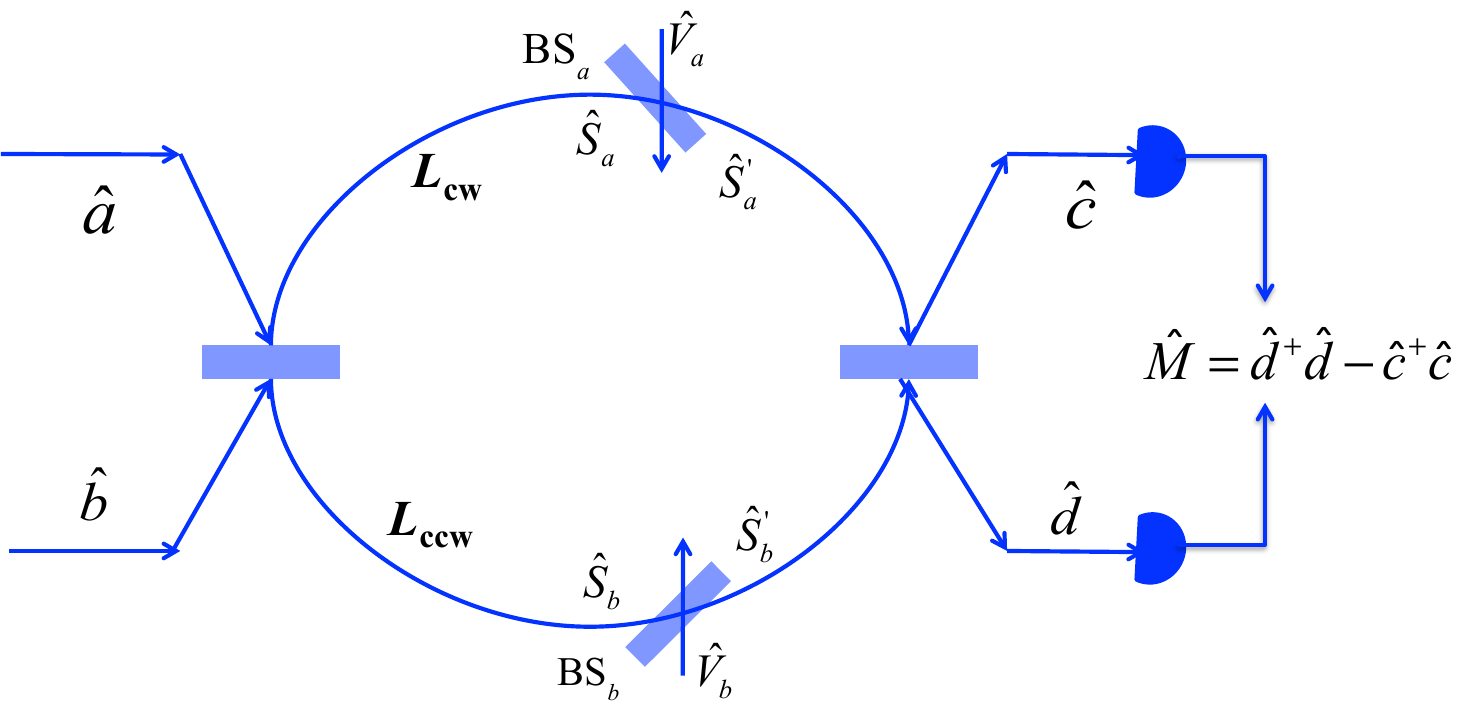}
\caption{The Sagnac interferometer in presence of loss in light propagation process, where the loss is modeled by the fictitious beam splitters.}
\label{4}
\end{figure}

In practice, there must be loss arising from the propagation of the light in the gyroscope. There are various origins of this loss, such as the changes in polarization, the widening of the spectrum, the optical Kerr effect, and so on.  The effect of the loss due to the propagation may be modeled by fictitious beam splitters BS$_{a}$ and BS$_{b}$, placed in the two optical paths \cite{rubin2007loss,huver2008entangled}, as described in Fig. \ref{4}. $\hat{S}_{i}$ denotes the field operator of the beam in the gyroscope at the input port of the BS$_i$, and $\hat{S}'_{i}$ denotes the field operator at the output port of the BS$_i$, $(i=a,b)$. The relation between $\hat{S}_{i}$ and $\hat{S}'_{i}$ can be expressed as follows:
\begin{equation}
\hat{S}'_{i}=t_{i}\hat{S}_{i}+r_{i}\hat{V}_{i},
\end{equation}
where $\hat{V}_{i}$ is the field operator for the environment, which is assumed to be vacuum, and $r_{i}$ and $t_{i}$ are the reflection and the transmission coefficients for mode $i$, which are assumed to be real. Clearly, $r^{2}_{i}=1-t^{2}_{i}$ stands for the loss in the mode $i$.

As mentioned in the preceding section, under the condition $\alpha =\beta $, the phase sensitivity is enhanced compared with the conventional gyroscope by a factor of $1/G$. When the loss is taken into account, following the argument in Sec. \ref{og}, we find the variance of the phase shift $\varphi $ is given by
\begin{eqnarray}
\Delta \varphi ^{2}&=&\frac{1}{2n^{2}G^{2}}+\frac{t_{a}^{2}r_{b}^{2}+t_{b}^{2}r_{a}^{2}}{4t_{a}^{2}t_{b}^{2}n^{2}G}\nonumber \\
&=&\left(\frac{1}{G}+\frac{t_{a}^{2}r_{b}^{2}+t_{b}^{2}r_{a}^{2}}{2t_{a}^{2}t_{b}^{2}}\right)\Delta \varphi ^{2}_{\mathrm{SN}},
\label{f}
\end{eqnarray}
where $\Delta \varphi _{\mathrm{SN}}$ is the shot-noise limit obtained for the ideal, no-loss condition.The phase sensitivity decreases due to the loss in the propagation. In the case $\theta =\pi +2\gamma $, the enhancement factor of the phase variance $R$ becomes
\begin{equation}
R=e^{-2r}+\frac{t_{a}^{2}r_{b}^{2}+t_{b}^{2}r_{a}^{2}}{2t_{a}^{2}t_{b}^{2}}
\label{tatb}
\end{equation}
When the transmission coefficients $t_{a}$ and $t_{b}$ approximate to $0$, the enhancement factor $R$ is going to be infinity, because, under this situation, no signal can be detected, and no information of the phase can be obtained either. 

\begin{figure}[t]
\centering
\begin{minipage}[b]{\linewidth}
\centering
\includegraphics[width=3.4in]{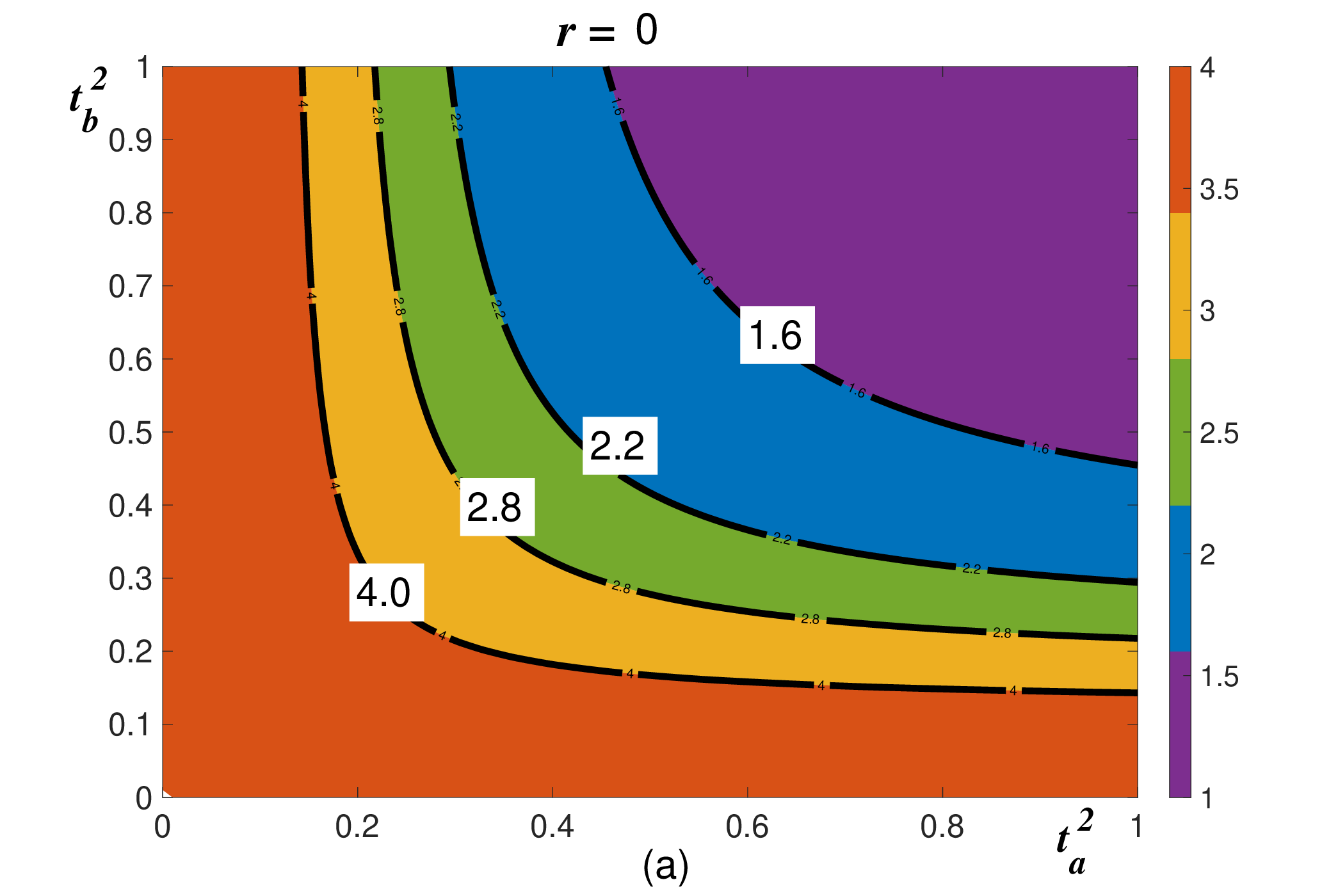}
%\caption{fig2}
\end{minipage}%
 \quad
\begin{minipage}[b]{\linewidth}
\centering
\includegraphics[width=3.4in]{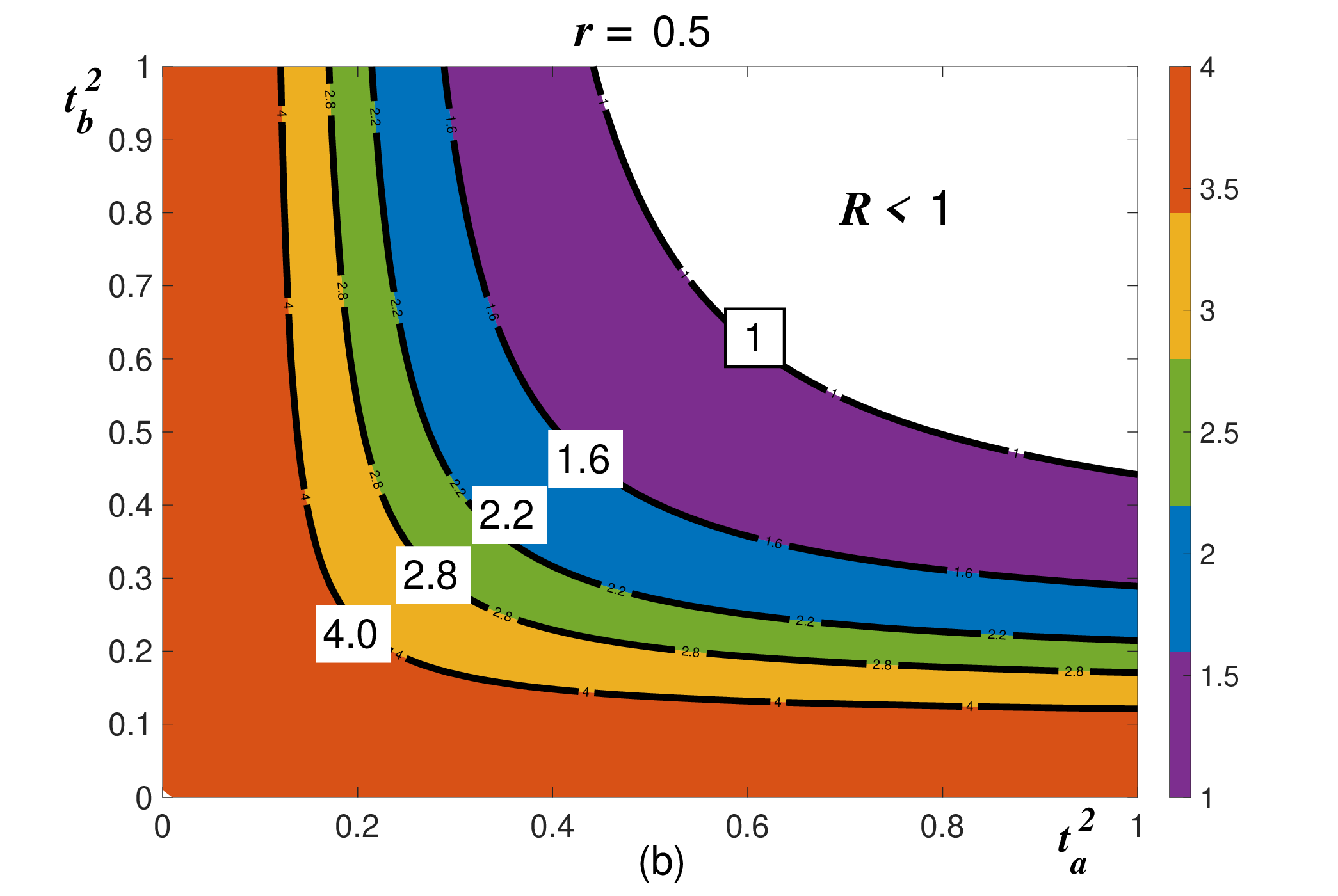}
%\caption{fig2}
\end{minipage}%
 \quad
 \begin{minipage}[b]{\linewidth}
\centering
\includegraphics[width=3.4in]{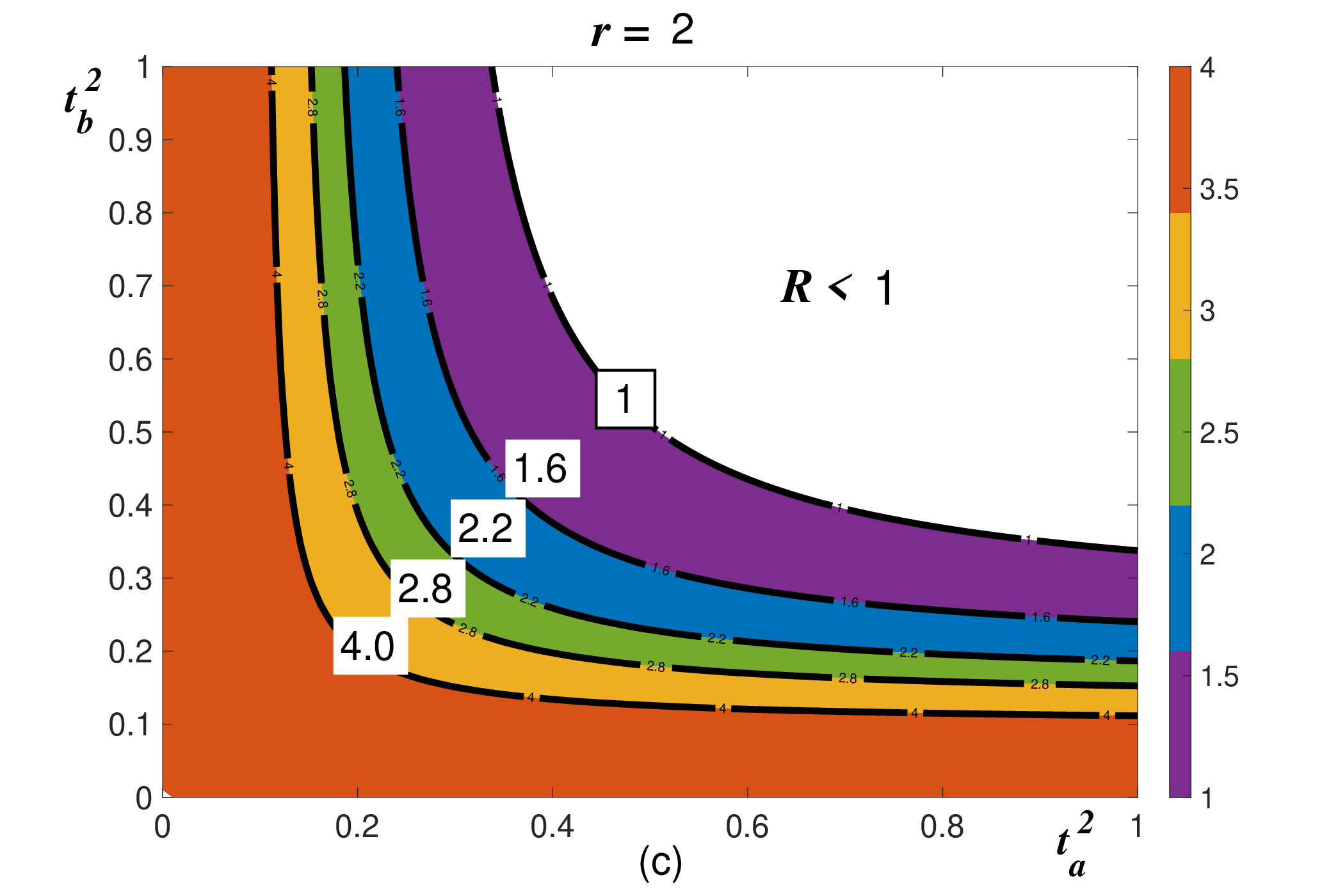}
%\caption{fig2}
\end{minipage}
\caption{Contour plot of the enhancement factor $R$ varying with the transmission coefficients $t^{2}_{a}$ and $t^{2}_{b}$. The area for $R <1$ (white colored) corresponds to the situation in which the phase sensitivity is below the shot-noise limit.}
\label{5}
\end{figure}

Figure \ref{5} shows the factor $R$ as a function of the transmission coefficients $t^{2}_{a}$ and $t^{2}_{b}$ for various values of the parametric gain coefficients $r$. It is clear that the area where $R<1$ corresponds to the situation where the phase sensitivity is below the shot-noise limit. When the parametric gain coefficient $r$ is fixed, the bigger the transmission coefficients $t^2_{a}$ and $t^2_{b}$ are, the smaller the ratio $R$ is, and the better the precision of the phase measurement is. Meanwhile, for given $t^2_{a}$ and $t^2_{b}$, the area where $R<1$ increases with the increase in the coefficient $r$, and the phase sensitivity is improved. That means the uncertainty of the phase measurement caused by the loss in light paths $L_{\rm{cw}}$ and $L_{\rm{ccw}}$ can be compensated by increasing the parametric gain coefficient $r$.

The reciprocity of the wave propagation in two directions plays an important role in the accuracy of the optical gyroscope. Let us consider the influence of the loss coming from the light propagation on the sub-shot-noise phase sensitivity under the conditions that the optical paths are reciprocal ($t^2_a =t^2_b $) and nonreciprocal ($t^2_a \neq t^2_b $), respectively.

\begin{enumerate}[(i)]

\item $t^2_{a}=t^2_{b}=T$

In general, the two optical paths should be reciprocal when the optical gyroscope is stationary. This means the losses in the two paths should be equal, thus we can set $t^2_{a}=t^2_{b}=T$. At this stage, the variance of the phase signal will be
\begin{equation}
\Delta \varphi ^{2}=\left(\frac{1}{G}+\frac{1-T}{T}\right)\Delta \varphi ^{2}_{\rm{SN}}.
\label{31}
\end{equation}
and Eq. (\ref{tatb}) is rewritten as
\begin{equation}
R=e^{-2r}+\frac{1-T}{T}
\end{equation}
It is easy to show that to obtain the sub-shot-noise limit the transmissivity needs to be 
\begin{equation}
T>\frac{1}{2-e^{-2r}}>\frac{1}{2}
\end{equation}
The above equation implies that it is always possible to obtain the sub-shot-noise limit if the loss in the optical path is no more than $3$dB.

\item $t^2_{a}\neq t^2_{b}$

In many applications, the losses in the two optical paths are most likely unequal, i.e. $t^2_{a}\neq t^2_{b}$. Suppose $t^2_{a}/ t^2_{b}=\kappa$ and $t^2_{b}=T$, such that $0<\kappa \leqslant 1$. The parameter $\kappa $ reflects the degree of the reciprocity, the two paths tend to be reciprocal as the parameter $\kappa $ approaches one, and vice versa. According to Eq. (\ref{f}), we have
\begin{equation}
\Delta \varphi ^{2}=\left(\frac{1}{G}+\frac{1}{2T}+\frac{1}{2\kappa T}-1\right)\Delta \varphi ^{2}_{\mathrm{SN}},
\end{equation}
and in the case of $\theta =\pi +2\gamma $, the ratio between $\Delta \varphi ^{2}$ and $\Delta \varphi ^{2}_{\rm{SN}}$ takes the following form
\begin{equation}
R=e^{-2r}+\frac{1}{2T}+\frac{1}{2\kappa T}-1.
\end{equation}
The required value of $\kappa $ for sub-shot-noise phase sensitivity ($R < 1$), of course, depends on $T$. From the above equation, it is always possible to achieve the sub-shot-noise limit measurement if $\kappa >1/3$ for $T=1$, or $\kappa >1/2$ for $T=3/4$.

\end{enumerate}

If we were to have a single-port input, by adjusting transmission coefficient of the BS, we would be able to compensate the unbalance of the loss. In our scheme, we have a two-port input before the first beam splitter. And the coherent beams will experience the OPA and the beam splitter (BS), and then go through the two paths of the interferometer. That means the amplitudes of the initial two coherent beams are not equal to those of the beams passing through the two arms. To overcome the unbalance between two paths by using the initial coherent states of different amplitudes, it is necessary to know the degree of the non-reciprocity between the two paths, which we define as $t^2_{a}/ t^2_{b}=\kappa$ here, and then adjust the parameters of OPA and  BS to ensure the amplitudes of the lights going through the two arms are balanced. On the other hand, there are many factors that can induce the non-reciprocity, such as the thermal fluctuation, light scattering, or other disturbing from outside. The degree of the non-reciprocity usually is unknown in practice. Therefore, optimizing the two amplitudes of the initial coherent states could be possible via an adaptive control with feedback.

In addition to the loss in propagation, the photon loss can occur in the detection, and alters the phase sensitivity of an interferometer sensor as well \cite{olivares2007optimized,sparaciari2016gaussian}. The effect of the photon loss in the detection is described in Appendix A. Comparing  Eq. (\ref{31}) with Eq. (\ref{A11}), we notice that  the loss in the detection and the loss in the propagation have the same effect on the phase sensitivity. After all, the final result is to shrink the number of photons to be detected no matter where the loss occurs (in the propagation or in the detection). From that point of view, the loss we discussed in this section can be regarded as the one that occurs not only in the propagation but also in the detection. \textcolor{magenta}{If the two kinds of loss are considered at the same time, there will be a multiplication of the two factors; $T$ is to be replaced by $T \tau $, where $\tau$ is the detection efficiency, shown in (A11).}

\section{Quantum Cr\'{a}mer-Rao Bound}
\label{qcrb}

For a probing system, the limit of the accuracy of the parameter estimation is given by its quantum Cr\'{a}mer-Rao Bound (QCRB), which is independent of the measurement strategy, and the limit may not always be achieved by any kind of measurement. In this section, we investigate the phase sensitivity for the gyroscope system to show whether the quantum Fisher information is saturated by intensity difference measurement. Unlike the numerical result reported in Ref. \cite{birrittella2015coherently}, here we will provide an analytical expression of the QCRB of the two-mode squeezed coherent state.

According to the discussion in Sec. \ref{og}, for the present measurement strategy the optimal phase sensitivity, which is  $\Delta \varphi ^{2} =1/[2n(e^{2r})^{2}]$, is achieved under the condition that the complex amplitudes of the two initial coherent beams are the same, i.e. $\alpha =\beta =\eta =\sqrt{n}e^{\rm{i}\gamma}$, and the phase match $\theta= \pi +2\gamma$ is satisfied (see Case I in Sec. \ref{og}). Therefore, the state of the light field injected into the gyroscope can be expressed as $\left\vert \Psi \right\rangle = \hat{S} \left\vert \eta , \eta \right\rangle$. And the gyroscope can be represented by a unitary transformation $\hat{U}_{\rm{gyro}}$ in terms $\hat {a}$ and $\hat{b}$ (see Appendix C):
\begin{equation}
\hat{U}_{\rm{gyro}}=e^{\frac{1}{2} \varphi (\hat{a}^{\dag}\hat{b}-\hat{b}^{\dag}\hat{a})}=e^{-\mathrm{i}\varphi \hat{G}_{\rm{gyro}}},
\end{equation}
where $\hat{G}_{\rm{gyro}} =\frac{\mathrm{i}}{2}(\hat{a}^{\dag}\hat{b}-\hat{b}^{\dag}\hat{a})$. Therefore, the quantum Fisher information is $4\Delta G^{2}_{\rm{gyro}} $, and the QCRB of the system is given by \cite{paris2009quantum}
\begin{equation}
\Delta \varphi ^{2}_{\rm{QCRB}}=\frac{1}{4\Delta \hat{G}_{\rm{gyro}}^{2} }= \frac{1}{2n(e^{2r})^{2} + \sinh^{2}2r}.
\end{equation}
The phase sensitivity obtained by intensity-difference measurement will be equal to the QCRB in the case that the average photon number of the initial coherent field $\langle \hat{N} \rangle _{\mathrm{coh}}$ is far greater than that of TMSV field $\langle \hat{N} \rangle _{\mathrm{TMSV}}$, i.e. $2n \gg 2\sinh^{2} r$. As shown in Fig. \ref{6}, the curves of the phase sensitivity and the QCRB tend to overlap with the increase in the average photon number of each initial coherent beam $n$ and the average total photon number at the output port of the gyroscope $\langle \hat{N}\rangle _{\mathrm{tot}}$. This indicates that the intensity difference measurement is capable of saturating the QCRB under the condition $n \gg \sinh^{2} r$.

\begin{figure}[t]
\centering
\begin{minipage}[b]{\linewidth}
\centering
\includegraphics[width=3.4in]{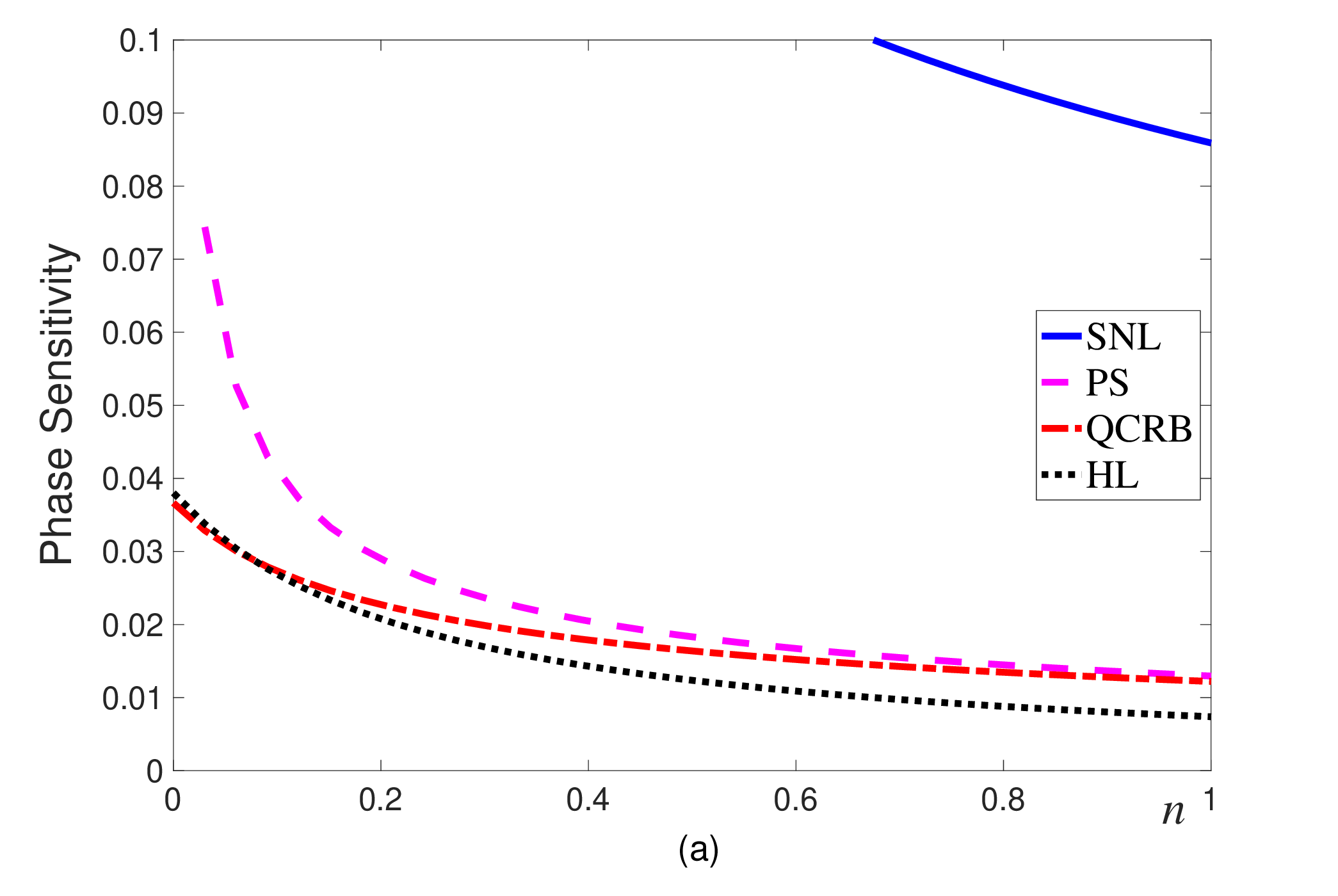}
%\caption{fig2}
\end{minipage}%
 \quad
\begin{minipage}[b]{\linewidth}
\centering
\includegraphics[width=3.4in]{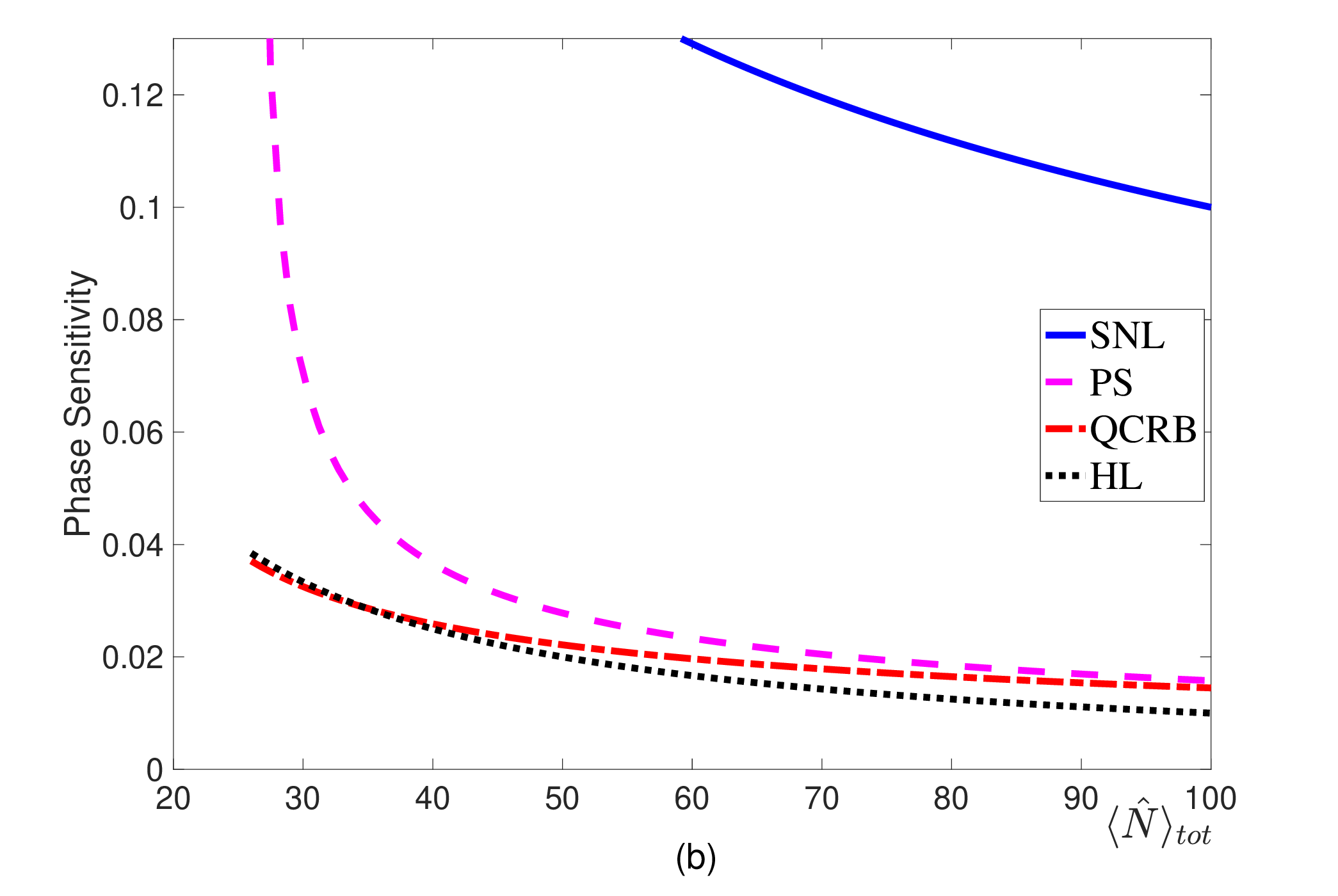}
%\caption{fig2}
\end{minipage}
\caption{The phase sensitivity $\Delta \varphi$ varying with (a) the average photon number of each one of the initial coherent beams $n$, and (b) the average total photon number of  the output light fields $\langle \hat{N} \rangle _{\mathrm{tot}}$. The parametric gain coefficient $r$ is set to be $2$, that $\langle \hat{N} \rangle _{\mathrm{tot}}\approx 26.3$ if the input coherent beams are in the vacuum state. SNL: shot-noise limit; PS: phase sensitivity of our interferometer scheme given by Eq. (\ref{15}); QCRB: Quantum Cr\'{a}mer-Rao Bound; HL: Heisenberg limit}
\label{6}
\end{figure}

The average number of the total photons at the output port is $\langle \hat{N}\rangle _{\mathrm{tot}} =2nG+2\sinh^{2}r$, based on Eq. (\ref{6}). Thus the shot-noise limit and the Heisenberg limit are $\Delta \varphi _{\rm{SN}}=1/\sqrt{2nG+2\sinh^{2}r}$ and $\Delta \varphi _{\rm{HL}}=1/(2nG+2\sinh^{2}r)$, respectively, which are also shown in Fig. \ref{6}. When the average photon number of the initial coherent field $n=0$, that means the light field used \textcolor{magenta}{as the probe} is the two-mode squeezed vacuum (TMSV) state. The total number of the photons exiting the gyroscope will then be $\langle \hat{N}\rangle _{\mathrm{tot}}=\langle \hat{N}\rangle _{\mathrm{TMSV}}=2\sinh^{2}r$, and the QCRB is 
\begin{equation}
\Delta \varphi ^{2}_{\rm{QCRB}}=\frac{1}{\langle \hat{N} \rangle _{\mathrm{tot}}(\langle \hat{N} \rangle _{\mathrm{tot}}+2)},
\end{equation}
which is consistent with the result in Ref. \cite{anisimov2010quantum}. 
Note that the QCRB for TMSV state is below the Heisenberg limit, which indicates that the Heisenberg limit---given by the inverse of the total photon number---is not the ultimate limit of the phase variance \cite{yurke19862,hofmann2009all}. As the phase sensitivity with intensity difference measurement becomes unbound, the intensity difference measurement strategy is no longer a good measurement scheme when the probing light field is in the TMSV state [see Eq. (14)]. In this case, the signal obtained by intensity difference measurement does not contain any information of the phase shift [see Eq. (\ref{c})]. \textcolor{magenta}{If losses were to be included, the QCRB can be calculated following, for example, Ref. \cite{demkowicz2009quantum}. Calculation of QCRB including losses, however,  becomes significantly more cumbersome and complex. }

\section{discussion}
\label{discussion}

Compared with the conventional gyroscope, a distinct advantage of our proposed scheme is the enhancement with a factor of $e^{2r}$ in the phase sensitivity $\Delta \varphi ^2$. As a consequence, the minimal detectable angular velocity will decrease with the factor $e^{-r}$, based on Eq. (\ref{a}). In practice, the minimum detectable angular velocity for a conventional fiber gyroscope is given as \cite{lefevre, wang}
\begin{eqnarray}
\Omega_{\rm{min}}^{\rm{cg}} & = & \frac{\lambda c}{2\pi DL}\frac{\sqrt{2e(1+\cos\varphi _{\rm{b}})\Delta f}}{\sqrt{R_DP}\sin\varphi _{\rm{b}}}\nonumber\\
& = &\frac{\lambda c}{2\pi DL}\sqrt{\frac{2e\Delta f\lambda}{R_D h c}}\frac{1}{\sqrt{2}\sin(\varphi _b/2)}\Delta \varphi _{\mathrm{SN}}, 
\end {eqnarray}
where $L$ is the length of the fiber, coil diameter $D=0.2\mathrm{m}$, $P$ is the output power at the detectors,  $\lambda $ is the wavelength, $R_{D}$ is the detector responsiveness which measures the input-output gain of a detector system (in the unit of A/W), $\phi_{b}$ is the bias phase, $e$ is the electron charge, $h$ is the Planck's constant, $c$ is the velocity of light and  $\Delta f$ is the count bandwidth of the output signal. If the the two-mode squeezed coherent state is adopted, as discussed in Sec. \ref{og}, according to Eq. (\ref{15}), the minimum detectable angular velocity can be expressed as
\begin{equation}
\Omega_{\rm{min}}^{\rm{qg}}  = \frac{\lambda c}{2\pi DL}\sqrt{\frac{2e\Delta f\lambda}{R_Dhc}}\Delta \varphi .
\end {equation}
It is obvious that the minimum detectable angular velocity decreases by a factor of $\sqrt {2}e^{-r}\sin (\phi_{b}/2)$. Note that for a fair comparison, we assume the total number of photons is the same for both the conventional gyroscope and the quantum gyroscope. 

For example, for a conventional fiber gyroscope with the parameters $L=5\mathrm{km}$, $D=0.2\mathrm{m}$, $P=250\mathrm{\mu W}$, $\lambda =1550\mathrm{nm}$, $R_{D}=1\mathrm{A/W}$, $\phi_{b}=3\pi /4$, and $\Delta f=0.01\mathrm{Hz}$, the minimum detectable angular velocity $\Omega ^{\rm{cg}}_{\rm{min}}$ is $4.2\times 10^{-5}\rm{(^{\circ })/h}$. Whereas, for the present gyroscope with the same parameters, the minimum detectable angular velocity $\Omega _{\rm{min}}^{\rm{qg}}$ can reach $5.5\times 10^{-6}\mathrm{(^{\circ })/h}$ in the case of $r=2.25$, which can be realized with the current technology \cite{agafonov2010two}. Besides, if the interferometer is injected with the light from the TMSV light field, as suggested in Ref. \cite{steuernagel2004approaching}, to realize a phase sensitivity equivalent to $5.5\times 10^{-6}\mathrm{(^{\circ })/h}$, the demand on the parametric gain $r$ would be no less than $9$, which is hard to achieve in practice. That is the reason why we adopt the output field of the OPA with the coherent light input as the light source of the gyroscope.

Typically, for the conventional optical gyroscope, the maximal sensitivity occurs at the point $\varphi =\pi /2$, and a bias phase $\phi_{b}$ is necessary to obtain the maximal sensitivity at the point $\varphi =0$. For our gyroscope scheme, according to Eq. (\ref{c}), when $\vert\alpha \vert =\vert \beta \vert$ the output signal is proportional to $\sin\varphi $, i. e. $\langle \hat{M} \rangle \propto \sin \varphi $. Thus $\partial \langle \hat{M} \rangle /\partial \varphi \propto \cos \varphi $, even if $r=0$. That shows the maximal sensitivity can be achieved at the working point $\varphi =0$ without the bias phase for our scheme. In turn, it indicates that the loss caused by the bias phase modulation in the conventional optical gyroscope can be avoided in our scheme. At the same time, our gyroscope scheme is robust against propagation loss for a wide range of the transmission coefficient $T$ and non-reciprocal $\kappa $. 

On the other hand, we can obtain the same phase sensitivity with a much smaller length of the light path, that is $\Omega ^{\mathrm{cg}}_{\mathrm{min}} =\Omega ^{\mathrm{qg}}_{\mathrm{min}} $, when $L^{\mathrm{qg}}=e^{-r}L^{\mathrm{cg}}$. As a result, the loss due to the light propagation in the fiber will be reduced due to the smaller length. Hence, any kind of loss that is proportional to the length of the fiber, such as loss coming from backscattering, the Kerr effect, the Faraday magnetic-optical effect etc., will decrease by a factor of $e^{-r}$. With a simple numeral calculation, we find these kinds of loss in our gyroscope will decrease to $10\%$ of those in the conventional one when we take the parametric gain of OPA $r=2.25$. 

In addition, several nonlinear interferometer schemes showing better performance than linear interferometers have been reported in recent years \cite{plick2010coherent,ou2012enhancement}. In the present paper, the uncertainty of the phase measurement can be expressed as $\Delta \varphi =e^{-2r}/\sqrt{\langle \hat{N} \rangle _{\mathrm{coh}}}$ by taking Eq.(\ref{e}) when $\theta =\pi +2\gamma $, which is comparable to the nonlinear interferometer in Ref. \cite{plick2010coherent}, with $\Delta \varphi =e^{-2r}/\sqrt{2\langle \hat{N} \rangle _{\mathrm{coh}}}$ in high gain limit. The output detection adopted in Ref. \cite{plick2010coherent}, however, is the intensity-sum measurement. Our gyroscope scheme using the intensity-difference measurement might well have advantages in that various technical noises can be eliminated with differencing. Also unlike the result in Ref. \cite{ou2012enhancement}, the influence of the loss in propagation can be compensated by enhancing the parametric gain of the OPA in our scheme.

\section{Conclusion}
\label{conclusion}

In conclusion, we have analyzed a new type of gyroscope with a coherent-boosted two-mode squeezed light field, and showed its sub-shot noise phase sensitivity. The phase sensitivity has been examined with different coherent light fields, and the results show that the optimal sensitivity occurs when the complex amplitudes of the two modes of the coherent light are equal to each other. The corresponding minimum detectable angular velocity has an enhancement with a factor of $e^{-r}$ compared with the conventional gyroscope with the same number of photons at the output---that factor would be $e^{-2r}$ if the same number of photons at the input is assumed. The effect of the loss from the light propagating through the interferometer is another significant factor in high precision measurement. 

The good news here is that it does not limit the enhancement of the measurement sensitivity, since it can also be offset by the increase of the parametric gain coefficient $r$. It is still possible to achieve sub-shot noise even in the noisy environment. 
In addition, we found that the intensity difference estimation strategy has the capability to reach Cr\'{a}mer-Rao Bound. It should be pointed out that the technologies involved in our scheme, such as the OPA and the intensity difference measurement, are highly accessible and well-established techniques. We also notice that the use of the orbital angular momentum states of light has been demonstrated for measuring rotation angle or vibrations \cite{liu2018squeezing,xiao2018orbital}. By contrast, our scheme is based on current gyroscope technique and can be integrated with existing optical gyroscope system. Therefore, we believe that our gyroscope scheme is positively suitable for practical application of quantum technique, and can be realized in near future.

\begin{acknowledgments}
This work was completed shortly after the untimely death of one of the authors, Jonathan P. Dowling. It is with immense gratitude that we recognize his friendship, his guidance, and his insight, which we deeply miss. This work was supported by the U.S. Army Research Office with the grant W911NF-17-1-0541, as well as the U.S. Air Force Office of Scientific Research,  the National Science Foundation, and the Program of Aboard Learning for Faculty Development in Universities of Shanghai China.
\end{acknowledgments}

\appendix
\section{The variance of the phase shift}

Assume the quantum efficiency of the two detectors are the same and equal to $\tau$ (where $\tau \leq 1$), the operators of the intensity detection are $\hat{m}_{d}=\tau \hat{d}^{\dagger}\hat{d}$ and $\hat{m}_{c}=\tau \hat{c}^{\dagger}\hat{c}$, respectively. In this condition, the operator for intensity difference is given as \cite{agliati2005quantum}
\begin{eqnarray}
\hat{M}_{\tau} & = & \tau \hat{d}^{\dagger}\hat{d}-\tau \hat{c}^{\dagger}\hat{c}=\tau\hat{M}\\
\widehat{M_{\tau} ^{2}} & = & \tau ^2 (\hat{d}^{\dagger}\hat{d}-\hat{c}^{\dagger}\hat{c})^2 +\tau (1-\tau)(\hat{d}^{\dagger}\hat{d}+\hat{c}^{\dagger}\hat{c})\nonumber \\
& = & \tau ^2 \hat{M}^2 +\tau (1-\tau)(\hat{d}^{\dagger}\hat{d}+\hat{c}^{\dagger}\hat{c})
\end{eqnarray}
And the variance on the intensity difference measurement is 
\begin{eqnarray}
\Delta M_{\tau}^2 & = & \langle \widehat{M_{\tau} ^{2}}\rangle -\langle \hat{M}_{\tau}\rangle^{2} \nonumber \\
& = & \tau ^{2}\langle \hat{M}^2 \rangle +\tau (1-\tau )\langle N \rangle _{tot}- \tau ^{2}\langle \hat{M}\rangle ^{2} \nonumber \\
& = & \tau ^2\Delta M^2+\tau (1-\tau)\langle \hat{N} \rangle _{tot},
\end{eqnarray}
where $\Delta M^2 = \langle \hat{M}^2 \rangle - \langle \hat{M}\rangle ^{2} $. Then, with the method of linear error propagation the phase sensitivity is
\begin{equation}
\Delta \varphi ^2 =\frac{\Delta M^2_{\tau}}{\vert \partial\langle M_{\tau}\rangle /\partial \varphi \vert ^2}=\frac{\tau ^2\Delta M^2+\tau (1-\tau )\langle \hat{N} \rangle _{tot}}{\tau ^2 \vert \partial\langle M\rangle /\partial \varphi \vert ^2}
\label{A4}
\end{equation}

Firstly, we are concerned with the case of idea detection, that is $\tau =1$, then the uncertainty of the phase shift $\varphi$ takes the following form
\begin{equation}
\Delta \varphi ^2 =\frac{\Delta M^2}{\vert \partial\langle M\rangle /\partial \varphi \vert ^2},
\label{A5}
\end{equation}
which is the Eq. (\ref{10}) in the Section \ref{og}. According to Eq. (\ref{8}) , there is 
\begin{eqnarray}
\Delta M^2 & = & \langle \hat{M}^2 \rangle - \langle \hat{M}\rangle ^{2} \nonumber \\
& = &\left\langle [ (\hat{a}^{\dag }\hat{a}-\hat{b}^{\dag }\hat{b})\cos \varphi +(\hat{a}^{\dag }\hat{b}+\hat{a}\hat{b}^{\dag })\sin \varphi ]^{2} \right \rangle \nonumber \\
& - &  \langle (\hat{a}^{\dag }\hat{a}-\hat{b}^{\dag }\hat{b})\cos \varphi +(\hat{a}^{\dag }\hat{b}+\hat{a}\hat{b}^{\dag })\sin \varphi \rangle ^{2}. 
\end{eqnarray}
Meanwhile  according to Eq. (\ref{c}), we have
\begin{eqnarray}
\frac{\partial \langle \hat{M}\rangle}{\partial \varphi} &=&-(\left\vert \alpha \right\vert ^{2}-\left\vert \beta \right\vert ^{2})\sin \varphi +\{(\alpha ^{\ast }\beta +\alpha \beta ^{\ast})\cosh 2r  \nonumber \\
& &-\frac{1}{2}[e^{\mathrm{i}\theta }(\alpha ^{\ast 2}+\beta ^{\ast 2})+e^{-\mathrm{i}\theta }(\alpha ^{2}+\beta ^{2})]\sinh 2r\}   \nonumber\\
& &\times \cos \varphi .
\end{eqnarray}
Since we focuse on the working point at $\varphi =0$, in the above two  equations the terms containing $\sin \varphi $ can be neglected.
\begin{eqnarray}
\Delta M^2 & = &\left\langle (\hat{a}^{\dag }\hat{a}-\hat{b}^{\dag }\hat{b})^{2}\right\rangle - \langle (\hat{a}^{\dag }\hat{a}-\hat{b}^{\dag }\hat{b})\rangle ^{2} \nonumber \\
& & =\left\vert \alpha \right\vert ^2 + \left\vert \beta \right\vert ^2\\
\frac{\partial \langle \hat{M}\rangle}{\partial \varphi} &=&\{(\alpha ^{\ast }\beta +\alpha \beta ^{\ast})\cosh 2r -\frac{1}{2}[e^{\mathrm{i}\theta }(\alpha ^{\ast 2}+\beta ^{\ast 2})\nonumber \\
& & +e^{-\mathrm{i}\theta }(\alpha ^{2}+\beta ^{2})]\sinh 2r\} ,
\end{eqnarray}
where Eqs. (\ref{2}) and (\ref{3}) are used. Bring the above two equations into the Eq. (\ref{A5}), we obtain the uncertainty of the phase shift with ideal detection:
\begin{widetext}
\begin{equation}
\Delta \varphi ^{2}=\frac{\left\vert \alpha \right\vert ^{2}+\left\vert \beta \right\vert ^{2}}{\left\vert (\alpha ^{\ast }\beta +\alpha \beta ^{\ast })\cosh 2r-\frac{1}{2}[e^{\mathrm{i}\theta }(\alpha ^{\ast 2}+\beta ^{\ast 2})+e^{-\mathrm{i}\theta }(\alpha ^{2}+\beta ^{2})]\sinh 2r\right\vert ^{2}},
\end{equation}
\end{widetext}
which is the Eq. (\ref{b}) in the Section \ref{og}.

Next, we turn to a briefly discussion about the effect of the quantum efficiency on the phase sensitivity. In the case $\alpha =\beta=\sqrt{n}e^{\mathrm{i}\gamma }$, based on Eq. (\ref{A4}) we have
\begin{eqnarray}
\Delta \varphi ^2  & = & \frac{\tau ^2 (2n)+\tau (1-\tau)(2nG)}{\tau ^2\vert 2nG\vert ^2} \nonumber \\
 & = & \left(\frac{1}{G}+\frac{1-\tau}{\tau }\right)\Delta \varphi ^2_{\mathrm{SN}}.
\label{A11}
\end{eqnarray}
Therefore, the the enhancement factor of the phase uncertainty becomes $R= \frac{1}{G}+\frac{1-\tau}{\tau }$. To make the phase uncertainty go below shot-noise limit, the condition $\tau >\frac{G}{2G-1}$ must be satisfied. For the optimal case of $\cos (\theta -2\gamma) = -1$, the quantum efficiency of the detectors $\tau >\frac{1}{2}$ is required.

\section{Noise Reduction Factor}

Much attention has been paid to the two-mode squeezed states due to the quantum correlation between its two modes. One of the characteristics of its quantum correlation is the suppression on the intensity difference noise between the two modes under certain conditions. 

Assume the annihilation operators of two light beams are $\hat{i}_{1}$ and $\hat{i}_{2}$, then the intensity difference between the two beams can be expressed as $\hat{I}_{\mathrm{diff}}=\hat{i}_{1}^{\dagger}\hat{i}_{1}- \hat{i}_{2}^{\dagger}\hat{i}_{2}$. In generally, for two independent coherent light beams, the state of which is $\left\vert  \chi _{1}  \right\rangle \left\vert\chi _{2} \right\rangle$, the variance of the intensity difference between the two beams is
\begin{eqnarray}
\Delta I_{\mathrm{diff}} & = & \langle (\hat{i}_{1}^{\dagger}\hat{i}_{1}- \hat{i}_{2}^{\dagger}\hat{i}_{2})^{2}\rangle - \langle \hat{i}_{1}^{\dagger}\hat{i}_{1}- \hat{i}_{2}^{\dagger}\hat{i}_{2}\rangle ^{2} \nonumber \\
& &=\vert \chi _{1}\vert ^{2} +\vert \chi _{2}\vert ^{2} .
\end{eqnarray}
This is the shot-noise limit of the intensity-difference between any two light beams, which is equal to the total average photon number of the two light beams. 

For the two-mode squeezed light beams, when we ignore the quantum correlation between the two modes, the uncertainty of the intensity difference should be equal to the total average photon number $\langle \hat{N} \rangle _{\mathrm{tot}}$. That is to say the the shot-noise limit in detecting the intensity difference between the two modes $\hat{a}$ and $\hat{b}$, according to Eq. (\ref{7}), is
\begin{eqnarray}
\Delta{k}_{\mathrm{SN}} &=& (\left\vert \alpha \right\vert ^{2}+\left\vert \beta \right\vert ^{2})\cosh 2r   \nonumber\\
& &-(\alpha ^{\ast }\beta ^{\ast }e^{i\theta }+\alpha \beta e^{-i\theta })\sinh 2r \nonumber \\
&=&  (\left\vert \alpha \right\vert ^{2}+\left\vert \beta \right\vert ^{2})\cosh 2r \nonumber \\ 
& &-2\vert \alpha \vert \vert \beta \vert \sinh 2r \cos(\theta -\gamma _{\alpha }-\gamma _{\beta })
\end{eqnarray}
in the case that $\alpha =\vert \alpha \vert e^{\mathrm{i}\gamma _{\alpha }}$ and $\beta =\vert \beta \vert e^{\mathrm{i}\gamma _{\beta }}$. However, for the two-mode squeezed state, i.e., $\hat{S} \left\vert  \alpha  \right\rangle \left\vert \beta \right\rangle$,  the variance of the intensity difference between the two modes $\hat{a}$ and $\hat{b}$ in fact is
\begin{eqnarray}
\Delta{k} & = &  \langle (\hat{a}^{\dagger}\hat{a}- \hat{b}^{\dagger}\hat{b})^{2}\rangle - \langle \hat{a}^{\dagger}\hat{a}- \hat{b}^{\dagger}\hat{b}\rangle ^{2} \nonumber \\
& = &\vert \alpha \vert ^{2} + \vert \beta \vert ^{2} 
\end{eqnarray}
The rate between the shot-noise limit $\Delta{k}_{\mathrm{SN}}$ and the variance $\Delta{k} $, defined as Noise Reduction Factor (NRF),  is given as \begin{widetext}
\begin{equation}
\mathrm{NRF}=\frac{\Delta \hat{k}^{2}}{\Delta \hat{k}_{\mathrm{SN}}^{2}} = \frac{\left\vert \alpha \right\vert ^{2}+\left\vert \beta \right\vert ^{2}}{[(\left\vert \alpha \right\vert ^{2}+\left\vert \beta \right\vert ^{2})\cosh 2r-2\left\vert \alpha \right\vert \left\vert \beta \right\vert \sinh 2r\cos (\theta -\gamma _{\alpha }-\gamma _{\beta})]}.
\label{NRF}
\end{equation}
\end{widetext}
Obviously, when $\mathrm{NRF} < 1$ the quantum noise is reduced, vice versa.

When $\alpha =\beta = \sqrt{n}e^{\mathrm{i}\gamma }$, there is
\begin{equation}
\mathrm{NRF}=\frac{1}{\cosh 2r -\sinh2r \cos (\theta -2\gamma)}=\frac{1}{G}.
\label{B5}
\end{equation}
It shows the quantum noise will decrease when the condition $G>1$ is satisfied. The relation described in Eq. (\ref{B5}) shows good agreement with the theoretic analysis and experimental results in Refs. \cite{aytur1990pulsed,adamyan2004continuous,bondani2007sub}, in which the reduction is attributed to the quantum correlation between the photon number of the two beams.

Based on Eq. (\ref{NRF}) the NRF for Case II in Section \ref{og} ($\left\vert \beta \right \vert / \left\vert \alpha \right\vert =\epsilon$, but $\gamma _{\alpha}=\gamma _{\beta } =\gamma $) is given as
\begin{equation}
\mathrm{NRF}=\frac{1+\epsilon ^2}{(1+\epsilon ^2)\cosh2r - 2\epsilon \sinh2r\cos(\theta -2\gamma)}
\label{B6}
\end{equation}
And there is 
\begin{equation}
\mathrm{NRF}=\frac{1}{\cosh 2r -\sinh2r \cos (\theta -2\gamma-\delta)}
\label{B7}
\end{equation}
 for Case III in Section \ref{og} ($\left\vert \alpha \right \vert =\left\vert \beta \right\vert =\sqrt{n}$, but $\gamma _{\beta} - \gamma _{\alpha} = \delta$).

\section{Calculation for Quantum Cr\'{a}mer-Rao Bound}

When a certain variable $\lambda$ (for example) is estimated, we usually use the Cramer-Rao inequality to show the limit of the mean square error of the variable $\lambda$, that is \textcolor{magenta}{$\Delta \lambda ^2\ge 1/pF(\lambda)$ (where $p$} is the is the number of measurements
and $F(\lambda)$ is the so-called Fisher Information). That means the limit of the variance, or the Cramer-Rao Bound, is \textcolor{magenta}{$\Delta \lambda ^{2}_{CRB} = 1/pF(\lambda)$}. 

Note that, the \textcolor{magenta}{probing} light field is in the two-mode squeezed state, and the state to be detected is a pure state. According to Ref. \cite{paris2009quantum}, the quantum Fisher information has a simple form $F(\lambda)=4\langle \psi _0 \vert \Delta G^2_{\mathrm{sys}} \vert \psi _0 \rangle $, where $\vert \psi _0 \rangle$ is the \textcolor{magenta}{probing} state and $\hat{G}_{\mathrm{sys}}$ is the corresponding Hermitian generator of the system. 

As described in Section \ref{qcrb}, the \textcolor{magenta}{probing} state of our gyroscope is $\left\vert \Psi \right\rangle = \hat{S} \left\vert \eta , \eta \right\rangle$, then the light field to be detected will be in the state $\left\vert \Psi (\varphi)\right\rangle = \hat{U}_{\rm{gyro}} \left\vert \Psi \right\rangle $, where $ \hat{U}_{\rm{gyro}}= \hat{U}^{-1}_{\rm{BS}}\hat{P}_{\varphi} \hat{U}_{\rm{BS}}$. In terms of the operators of the modes $\hat{a}$ and $\hat{b}$, there are \cite{jarzyna2012quantum}
\begin{eqnarray}
\hat{U}_{\mathrm{BS}}&=& e^{\mathrm{i}\frac{\pi}{4}(\hat{a}^{\dagger}b+\hat{b}^{\dagger}a)}, \\
\hat{P}_{\varphi }&=&e^{\mathrm{i}(-\frac{\varphi}{2}\hat{a}^{\dagger}\hat{a}+\frac{\varphi}{2}\hat{b}^{\dagger}\hat{b})}=e^{-\mathrm{i}\frac{\varphi}{2}(\hat{a}^{\dagger}\hat{a}-\hat{b}^{\dagger}\hat{b})},
\end{eqnarray}
\textcolor{magenta}{here we assume that the phase is imprinted symmetrically. That means the model with symmetrically distributed phase shift is used \cite{jarzyna2012quantum}.} We then get
\begin{equation}
\hat{U}_{\rm{gyro}}=e^{\frac{1}{2} \varphi (\hat{a}^{\dag}\hat{b}-\hat{b}^{\dag}\hat{a})}=e^{-\mathrm{i}\varphi \hat{G}_{\rm{gyro}}},
\end{equation}
where $\hat{G}_{\rm{gyro}} =\frac{\mathrm{i}}{2}(\hat{a}^{\dag}\hat{b}-\hat{b}^{\dag}\hat{a})$.

%The quantum Fisher information  is given by, according to Ref. \cite{paris2009quantum}
%\begin{equation}
%F(\varphi)=2\sum _{nm}\frac{\vert \langle\psi (\varphi)_{m}\vert \partial _{\varphi }\rho (\varphi )\vert \psi (\varphi)_{n}\rangle }{\rho _{n}+\rho _{n}}, 
%\end{equation}
%where$\partial _{\varphi }=\partial /\partial \varphi $, and $\rho (\varphi ) = \sum_{n}\rho _{n} \vert \psi (\varphi)_{n}\rangle\langle \psi (\varphi)_{n} \vert $ is the density matrix of the state $\vert \Psi (\varphi)\rangle $. Here we have
%\begin{equation}
%\partial _{\varphi }\rho (\varphi )=\mathrm{i} \hat{U}_{\rm{gyro}}[\hat{G}_{\rm{gyro}}, \rho _{0}]\hat{U}_{\rm{gyro}}^{\dagger},
%\end{equation}
%where $\rho _{0}$ is the density matrix of the state $\vert \Psi \rangle $. 

Note that the \textcolor{magenta}{probing} light field is in the two-mode squeezed state, and the state $\left\vert \Psi (\varphi)\right\rangle = \hat{U}_{\rm{gyro}} \left\vert \Psi \right\rangle $ we detected is a pure state. Therefore, according to Ref. \cite{paris2009quantum}, the quantum Fisher information can be written as
\begin{eqnarray}
F(\varphi)&=&4\langle \Psi \vert \Delta G^{2}_{\rm{gyro}} \vert \Psi \rangle \nonumber\\
&=& 4(\langle \eta , \eta  \vert \hat{S}^{\dagger} \hat{G}^{2}_{\rm{gyro}} \hat{S} \vert \eta , \eta \rangle-\langle \eta , \eta  \vert \hat{S}^{\dagger} \hat{G}_{\rm{gyro}} \hat{S} \vert \eta , \eta \rangle^{2})\nonumber \\
&=&2n(e^{2r})^{2}+\sinh^{2}2r,
\end{eqnarray}
where Eqs. (\ref{2}) and (\ref{3}) are used. The QCRB of the system is then obtained as
\begin{equation}
\Delta \varphi ^{2}_{\rm{QCRB}}=\frac{1}{4\Delta \hat{G}_{\rm{gyro}}^{2} }= \frac{1}{2n(e^{2r})^{2} + \sinh^{2}2r}.
\end{equation}

\providecommand{\noopsort}[1]{}\providecommand{\singleletter}[1]{#1}%

\end{document}